\documentclass[fleqn,usenatbib]{mnras}
\usepackage{newtxtext,newtxmath}

\usepackage[T1]{fontenc}

\DeclareRobustCommand{\VAN}[3]{#2}
\let\VANthebibliography\thebibliography
\def\thebibliography{\DeclareRobustCommand{\VAN}[3]{##3}\VANthebibliography}

\usepackage{graphicx}	
\usepackage{amsmath}	
\usepackage{pdflscape}

\newcommand{\astropy}{{\sc Astropy}}
\newcommand{\numpy}{{\sc numpy}}
\newcommand{\python}{{\sc python}}
\newcommand{\scipy}{{\sc scipy}}
\newcommand{\matplotlib}{{\sc matplotlib}}

\title[Variability of BCGs at high radio frequencies]{The variability of brightest cluster galaxies at high radio frequencies}

\author[Tom Rose et al.]{Tom Rose$^{1,2,3}$\thanks{E-mail: thomas.rose@uwaterloo.ca},
Alastair Edge$^{1}$, 
Sebastian Kiehlmann$^{4,5}$, 
Junhyun Baek$^{6}$,
Aeree Chung$^{6}$,  \newauthor
Tae-Hyun Jung$^{7}$,
Jae-Woo Kim$^{7}$,
Anthony C. S. Readhead$^{8}$,
and Aidan Sedgewick$^{1}$
\\
\noindent $^{1}$Centre for Extragalactic Astronomy, Durham University, DH1 3LE, UK\\
$^{2}$ Department of Physics and Astronomy, University of Waterloo, Waterloo, ON N2L 3G1, Canada \\
$^{3}$ Waterloo Centre for Astrophysics, Waterloo, ON N2L 3G1, Canada \\
$^{4}$ Institute of Astrophysics, Foundation for Research and Technology-Hellas, GR-71110 Heraklion, Greece \\ $^{5}$ Department of Physics, University of Crete, GR-70013 Heraklion, Greece \\
$^{6}$ Department of Astronomy, Yonsei University, 50 Yonsei-ro, Seodaemun-gu, Seoul 03722, Korea\\
$^{7}$ Korea Astronomy and Space Science Institute, 776 Daedeokdae-ro, Yuseong-gu, Daejeon 34055, Korea \\
$^{8}$ Owens Valley Radio Observatory, California Institute of Technology, Pasadena, CA 91125, USA \\
}

\date{Accepted XXX. Received YYY; in original form ZZZ}

\pubyear{2021}

\begin{document}
\label{firstpage}
\pagerange{\pageref{firstpage}--\pageref{lastpage}}
\maketitle

\begin{abstract}
Variability of a galaxy's core radio source can be a significant consequence of AGN accretion. However, this variability has not been well studied, particularly at high radio frequencies. As such, we report on a campaign monitoring the high radio frequency variability of 20 nearby, cool-core brightest cluster galaxies. From our representative sample, we show that most vary significantly over timescales of approximately one year and longer. Our highest cadence observations are at 15 GHz and are from the Owens Valley Radio Observatory (OVRO). They have a median time interval of 7 days and mostly span between 8 and 13 years. We apply a range of variability detection techniques to the sources' lightcurves to analyse changes on week to decade long timescales. Most notably, at least half of the sources show 20 per cent peak to trough variability on 3~year timescales, while at least a third vary by 60 per cent on 6~year timescales. Significant variability, which is important to studies of the Sunyaev-Zel'dovich Effect in the radio/sub-mm, is therefore a common feature of these sources. We also show how the variability relates to spectral properties at frequencies of up to 353 GHz using data from the Korean VLBI network (KVN), the NIKA2 instrument of the IRAM 30m telescope, and the SCUBA-2 instrument of the James Clerk Maxwell Telescope.
\end{abstract}


\begin{keywords}
galaxies: clusters -- galaxies: radio continuum
\end{keywords}



\section{Introduction}

\begin{table*}
	\centering
	\caption[OVRO BCG monitoring sample]{The brightest cluster galaxies monitored by the OVRO 40m Telescope and the main details of the observations. The data below are for the lightcurves after they have been masked according to the criteria given in \S\ref{sec:masking}, and have had their extended emission subtracted according to the process described in \S\ref{sec:extended_emission_correction}.
	\newline
	Header clarifications: `Obs. / day' is the total number of observations divided by the time range over which they were taken, and the `Detectable variability' is defined as the mean flux density error divided by the mean flux.}
	\begin{tabular}{lccccccccc}
\hline
Source  &  RA  &  Dec  &  Redshift  &  Start date  &  Timespan  &  Obs.  &  Mean 15 GHz flux  &  Detectable \\
  &   &   &   &  MJD (ISO)  &  / yrs  &  / day  &   density / mJy  &  variability / \% \\
\hline
3C286 - Calibrator  &  13:31:08.400  &  30:30:33.120  &  0.8493  &  54474 (2008-01-09) &  13.7  &  0.42  &  3437  &  1.2 \\
DR21 - Calibrator  &  20:39:01.200  &  42:19:32.880  &  0.0  &  54473 (2008-01-08) &  13.7  &  0.46  &  19079  &  1.0 \\
RXJ0132.6-0804  &  01:32:41.112  &  -8:04:05.880  &  0.1489  &  56352 (2013-03-01) &  8.5  &  0.08  &  102  &  4.5 \\
J0439+0520  &  04:39:02.256  &  05:20:43.080  &  0.208  &  54476 (2008-01-11) &  13.6  &  0.12  &  319  &  3.7 \\
A646  &  08:22:09.600  &  47:05:52.800  &  0.1268  &  56323 (2013-01-31) &  8.6  &  0.09  &  59  &  6.8 \\
4C55.16  &  08:34:54.960  &  55:34:22.080  &  0.2411  &  56342 (2013-02-19) &  8.6  &  0.11  &  519  &  4.1 \\
A1348  &  11:41:24.240  &  -12:16:37.560  &  0.1195  &  56336 (2013-02-13) &  8.6  &  0.07  &  44  &  9.4 \\
3C264/NGC3862  &  11:45:05.016  &  19:36:22.680  &  0.0217  &  58432 (2018-11-10) &  2.8  &  0.12  &  126  &  8.2 \\
A1644  &  12:57:11.592  &  -17:24:34.200  &  0.0475  &  58459 (2018-12-07) &  2.8  &  0.04  &  200  &  1.9 \\
J1350+0940  &  13:50:22.080  &  09:40:09.840  &  0.13  &  54476 (2008-01-11) &  13.6  &  0.1  &  131  &  7.4 \\
J1459-1810/S780  &  14:59:28.800  &  -18:10:45.120  &  0.2357  &  54910 (2009-03-20) &  12.4  &  0.08  &  134  &  6.7 \\
A2052  &  15:16:44.640  &  07:01:18.120  &  0.0351  &  56324 (2013-02-01) &  8.6  &  0.09  &  100  &  5.9 \\
A2055  &  15:18:46.560  &  06:13:58.080  &  0.1019  &  56319 (2013-01-27) &  8.6  &  0.09  &  36  &  15.5 \\
J1558-1409  &  15:58:21.840  &  -14:09:59.040  &  0.097  &  54477 (2008-01-12) &  13.6  &  0.09  &  152  &  7.5 \\
J1603+1554  &  16:03:38.160  &  15:54:02.160  &  0.109  &  54761 (2008-10-22) &  12.8  &  0.11  &  236  &  4.1 \\
J1727+5510/A2270  &  17:27:23.520  &  55:10:53.040  &  0.24747  &  54480 (2008-01-15) &  13.5  &  0.11  &  236  &  4.1 \\
Z8276  &  17:44:14.400  &  32:59:29.400  &  0.075  &  56326 (2013-02-03) &  8.6  &  0.08  &  76  &  5.8 \\
J1745+398  &  17:45:37.752  &  39:51:20.880  &  0.267  &  58428 (2018-11-06) &  2.8  &  0.07  &  39  &  6.6 \\
A2390  &  21:53:36.720  &  17:41:44.880  &  0.228  &  56340 (2013-02-17) &  8.5  &  0.12  &  49  &  9.0 \\
A2415  &  22:05:38.640  &  -5:35:33.720  &  0.0573  &  56327 (2013-02-04) &  8.6  &  0.1  &  132  &  2.9 \\
A2627  &  23:36:42.720  &  23:55:23.880  &  0.127  &  56348 (2013-02-25) &  8.5  &  0.11  &  41  &  10.6 \\
RXJ2341.1+0018  &  23:41:06.960  &  00:18:33.480  &  0.2768  &  56328 (2013-02-05) &  8.6  &  0.11  &  98  &  4.5 \\
\hline
	\end{tabular}
	\label{tab:observations}
\end{table*}

The accretion of matter onto an active galactic nucleus (AGN) is associated with compact radio emission from partially self-absorbed synchrotron jets \citep{Remillard2006}. These powerful outflows transport energy out to large distances, where they heavily influence their environment by preventing runaway gas cooling, reducing the cold gas content at the centres of galaxy clusters, and curtailing star formation \citep{Edge2001, SalomeCombes2003, Rafferty2008}.

Accretion onto AGN occurs intermittently, and leads to a continual drift between `radio loud' and `radio quiet' phases \citep[][]{Wilson1995, Kellermann2016}. For the wider galaxy and cluster, this produces alternating periods of energy injection and cooling, evidence for which can be seen in X-ray cavities inflated in the intracluster medium \citep{McNamara2000, Hlavacek-Larrondo2012}.

These radio loud and quiet phases are associated with periods of gas heating and cooling, and the two have apparently existed in a fine balance since the earliest times of cluster formation \citep{Pratt2010, McDonald2018}. As a result, the energy injected into clusters typically offsets any radiative gas cooling. However, though this is true over long time intervals, shorter periods of heating and cooling still occur as AGN pass through their loud and quiet phases.

The radio variability of AGN is a product of many underlying physical processes which connect the host galaxy to its supermassive black hole \citep[for a review see][]{Hardcastle2020}. These include the rate and manner of the black hole fuelling \citep{King2007, Gaspari2013}, spectral index changes due to interactions within the AGN, and evolving energy densities within radio jets \citep{Blandford2019}. Studying and quantifying variability can therefore inform us about these processes. For example, the variability of a source across several years can indicate the mechanisms through which it is fuelled. On the other hand, because a source is unable to vary on timescales less than its light crossing time, short-term variability can be used as an indicator of its physical size. 

Despite the importance of radio observations to the study of fuelling and feedback cycles in galaxy clusters, it is not well known how much change the AGN at the centre of the process experience during their radio loud phases. This is particularly true at higher radio frequencies of around 10 - 100 GHz, where there has been a deficit of research. 

Here we present results from an ongoing 15 GHz survey of 20 X-ray selected brightest cluster galaxies (BCGs), whose variability we examine on week to decade long timescales. This work is a continuation of \citet{Hogan2015}, and includes the addition of several more BCGs. All the monitored sources contain an active core, reside in cool-core clusters, and have a significant radio core component in their spectral energy distributions (SEDs). These properties suggest strong, ongoing feedback, and allow us to detect variability with short observations. The main purpose of this work is to study the amplitude and timescales of variability in the radio cores of these sources, which are dependent upon the short timescale accretion processes of the AGN. We also show how this variability relates to the sources' spectral properties at radio frequencies of up to 353 GHz.

The paper is laid out as follows. In Section \ref{sec:sample_and_observations} we present our sample and give details of the target selection and observations. In Section \ref{sec:lightcurves} we show the lightcurves of our targets and corresponding spectral indices. In Section \ref{sec:detection_techniques} we discuss variability detection and quantification parameters, and then apply them to the lightcurves in Section \ref{sec:application_of_variaibility_detection_parameters}. In Section \ref{sec:discussion} we discuss our results and finish by presenting conclusions in Section \ref{sec:conclusions}.

\section{Sample and Observations}
\label{sec:sample_and_observations}
The brightest, core-dominated radio sources in this study are drawn from the sample presented in \citet{Hogan2015} and  \citet{Hogan2015b}. They are hosted by BCGs in X-ray selected clusters from the {\sl ROSAT} All-Sky Survey in the BCS \citep{ebeling1998}, eBCS  \citep{ebeling2000}, and REFLEX \citep{Bohringer2004} samples.
The uniform X-ray flux limit from the full high Galactic latitude sky ($|b|>20^\circ$) 
ensures a representative sampling of host clusters irrespective of the radio properties
of the BCG. Only when the BCG hosts a strong BL~Lac nucleus contributing to the total X-ray flux of the cluster does our selection favour a system with a bright radio source \citep[see][for a full discussion of this]{Green2017}.

From the full {\sl ROSAT} sample of 726 clusters, we selected systems in which the BCG hosted a radio source brighter than 50~mJy (at the time of selection) at 15~GHz and were either known to be core-dominated \citep{Hogan2014} or had a radio SED consistent with a Giga-Hertz peaked spectrum (GPS).

In addition, we added two clusters missed in the original BCS selection because they fell below the X-ray flux limit, RXCJ1350.3+0940 ($z=0.13$) and RXCJ1603.5+1554 ($z=0.109$). These were mis-identified as being AGN-dominated due to their BCG containing a strong, flat-spectrum radio source. Four clusters with a powerful radio source in their BCG were also added: 4C$+$55.16 \citep{iwasawa1999}, A2270 \citep{green2016}, RGB1745$+$398/RXCJ1745.6+3951 \citep{Green2017} and RXCJ2341.1+0018 \citep{green2016}. Although the monitored sample now consists of 20 sources, it should be noted that because several of them are more recent additions, they have been observed over a shorter timespan. 

We have also obtained multi-visit Korean VLBI Network (KVN), IRAM-30m NIKA2, ALMA, and JCMT SCUBA-2 observations over the period 2015 to 2021 to determine the spectral variability of this sample in the 15 to 353~GHz range. Additional higher frequency data were obtained for a further 13 BCGs below the OVRO declination limit of -20$^\circ$, or which were too faint at 15~GHz. These results are presented in an online appendix for completeness (Tables \ref{tab:NIKA2_obs1} and \ref{tab:SCUBA2_obs1}). 

Details of the observations obtained with each observatory are provided in the following subsections.

\subsection{OVRO}
The Owens Valley Radio Observatory (OVRO), based in California, has carried out a 15 GHz monitoring campaign of more than 1500 radio sources since 2007 using its 40m Telescope. Most of its targets are blazar Fermi-LAT gamma-ray candidates, the full sample of which can be seen in \citet{Richards2011}. Table \ref{tab:observations} lists the subsample of 20 BCGs from this survey whose variability is analysed in this paper, and gives the main details of the observations.

Of the 20 BCGs, five have been included in the OVRO campaign from its beginning. In January 2013, 11 more were added, followed by several others at various times since. Observations of the entire OVRO sample (including the 20 BCGs this paper focuses on) were carried out each week, so the median time interval between observations is 7 days. The mean time interval is slightly higher at 10 days due to occasional poor weather, telescope maintenance, and other observing issues.

The OVRO 40m Telescope uses off-axis dual-beam optics and a cryogenic receiver with 2~GHz equivalent noise bandwidth centered at 15~GHz. The observations are conducted in an ON-ON fashion so that one of the beams is always pointed on the source. This double switching technique is used to remove gain fluctuations as well as atmospheric and ground contributions \citep{Readhead1989}.

The two beams were rapidly alternated using a Dicke switch until May 2014. In May 2014 a new pseudo-correlation receiver replaced the old receiver and a 180~degree phase switch has been used since. The change of the receiver resulted in a significant reduction in the observational noise that is reflected in the light curve uncertainty estimates.

Relative calibration is obtained with a temperature-stable noise diode to compensate for gain drifts. The primary flux density calibrator is 3C~286 with an assumed value of 3.44~Jy \citep{Baars1977}, DR21 is used as secondary calibrator source.
Details of the observation and data reduction schemes are given in \citet{Richards2011}.

\subsection{NIKA2}

Following the success of the IRAM-30m GISMO observations presented in \citet{Hogan2015b}, we obtained data with the NIKA2 bolometer camera on the IRAM-30m that covers both 2 and 1.2mm bands simultaneously \citep{Perotto2020}. These observations were over two pooled campaigns in 2019 and 2020 and maps were made with the resulting data to optimise for faint, unresolved sources. 

The flux density measurements obtained with NIKA2 are given in Table \ref{tab:NIKA2_obs1} of Appendix \ref{sec:NIKA2Appendix}. For each NIKA2 measurement, the table also provides the nearest OVRO flux, and the spectral indices calculated with these measurements.

\subsection{KVN}

As an additional constraint on the flux density of the core component of our sources we also obtained KVN observations at 22 and 43~GHz over 5 semesters between 2018 and 2020. The three telescope interferometry provided by the KVN delivers an angular resolution of 6~mas at 22~GHz and 3~mas at 43~GHz. Therefore the compact core component that will dominate any variability is preferentially detected and the more extended, non-variable emission is resolved out. The data were calibrated using \texttt{AIPS} following a standard reduction procedure, and imaged using \texttt{DIFMAP}. In order to push the detection limit down to ~50 mJy or less, we applied source phase referencing and/or frequency phase transfer techniques at both bands \citep[][]{Rioja2011, Algaba2015}. Our targets are mostly unresolved or slightly resolved within the KVN synthesized beam, hence we use the peak flux density in further analysis.

The flux density measurements obtained with KVN are given in Table \ref{tab:KVN_obs1} of Appendix \ref{sec:KVNAppendix}. The table also provides the spectral indices calculated with these measurements.

\subsection{JCMT SCUBA2}

JCMT SCUBA2 353~GHz photometry for all 20 OVRO targets and 7 southern or fainter sources was obtained over five semesters between 2019 and 2021. The data were reduced using the standard SMURF 
reduction package and the observatory derived calibrations. All the previous observations for the objects were reduced in the same way, so the photometry now supersedes that in \citet{Hogan2015b} and \citet{cheale2019}.

The flux density measurements obtained with SCUBA2 are given in Table \ref{tab:SCUBA2_obs1} of Appendix \ref{sec:SCUBA2Appendix}. For each measurement, the table also provides the nearest OVRO flux, and the spectral indices calculated with these measurements.

\subsection{ALMA}

For seven of the OVRO targets we have single epoch observations with ALMA at frequencies close to either CO(1-0) or CO(2-1) in the rest frame of the target -- from the absorption line survey of \citet{Rose2019b}, or observations of extended molecular emission \citep{russell2019}. These mm observations can be compared to the other high frequency data, and are listed in Appendix \ref{sec:ALMA_data}.

\section{Lightcurves}
\label{sec:lightcurves}
15 GHz lightcurves of the 20 BCGs, obtained with the OVRO 40m Telescope, are shown in Figs. \ref{fig:combined_lightcurves_1}, \ref{fig:combined_lightcurves_2}, \ref{fig:combined_lightcurves_3}, and \ref{fig:combined_lightcurves_4}. These have been masked according to the procedure outlined in \S\ref{sec:masking} and had their extended emission subtracted as described in \S\ref{sec:extended_emission_correction}. We also include the KVN 22 GHz flux densities for comparison. 

Spectral indices are shown below the lightcurves, calculated with (i) contemporaneous 22 and 43 GHz KVN observations, (ii) 15 GHz OVRO and 150 GHz SCUBA2 observations, (iii) 15 GHz OVRO and 353 GHz NIKA2 observations, and (iv) 15 GHz OVRO and \mbox{$\approx$ 100 GHz} ALMA observations. These pairs of observations are separated by less than 10 days in the majority of cases -- a time interval significantly shorter than that on which variability is seen in all of the sources (as is shown later in Section \ref{sec:application_of_variaibility_detection_parameters}).

\begin{figure*}
	\includegraphics[width=1\textwidth]{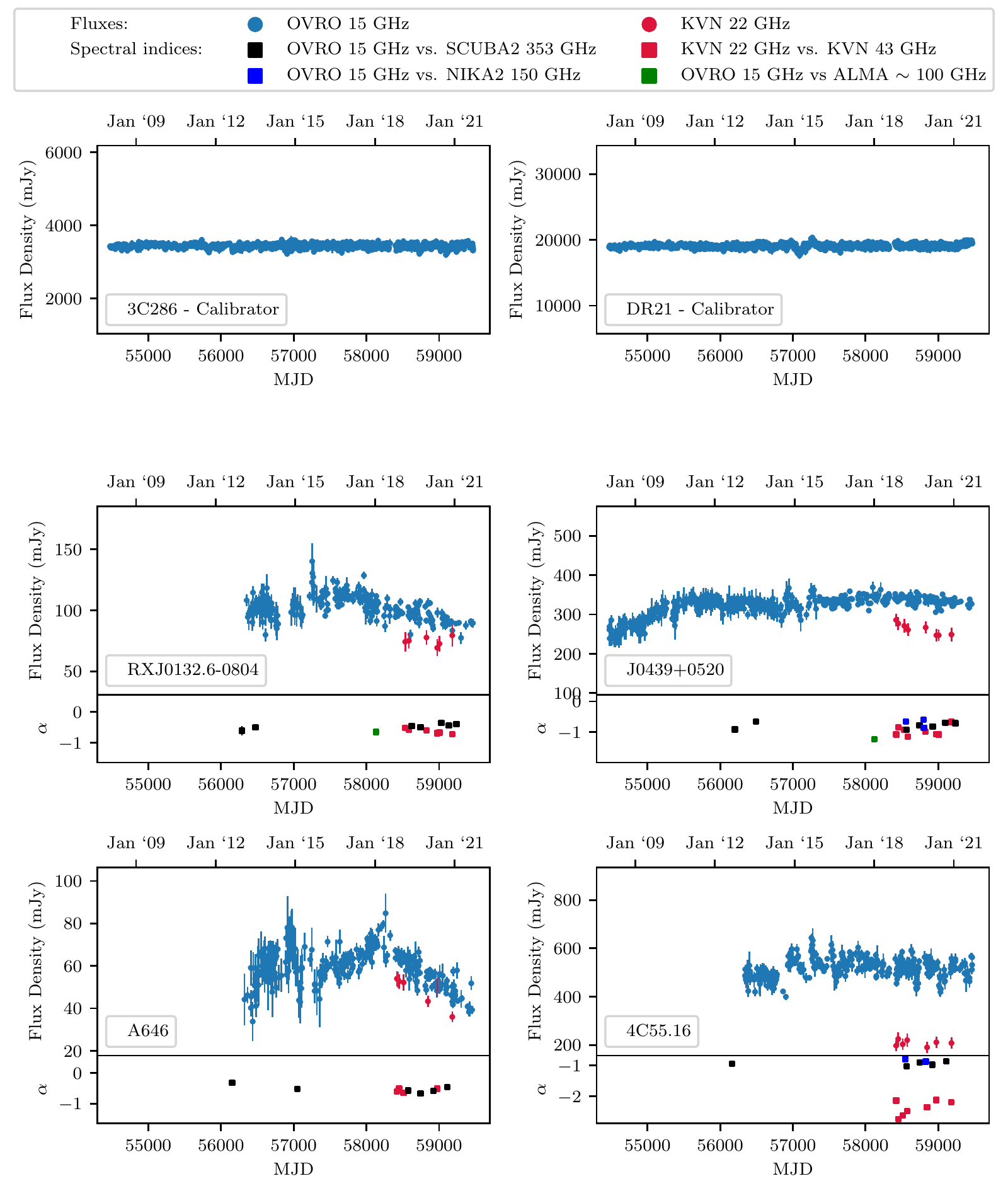}
     \caption[OVRO 15GHz lightcurves of BCGs]{15 GHz lightcurves (blue points) of our sample of cool-core BCGs obtained with the OVRO 40m telescope. The data has been masked according to the procedure given in \S\ref{sec:masking} and extended emission has been removed from several sources as described in \S\ref{sec:extended_emission_correction}. The Y-axes are scaled to each lightcurve's mean flux density to allow for easier comparison of the relative strength of the variability. The plots also include 22 GHz KVN flux densities (red points), and spectral indices calculated over the course of the observations. Continued in Figs. \ref{fig:combined_lightcurves_2}, \ref{fig:combined_lightcurves_3}, and \ref{fig:combined_lightcurves_4}.}
    \label{fig:combined_lightcurves_1}
\end{figure*}

\begin{figure*}
	\includegraphics[width=1\textwidth]{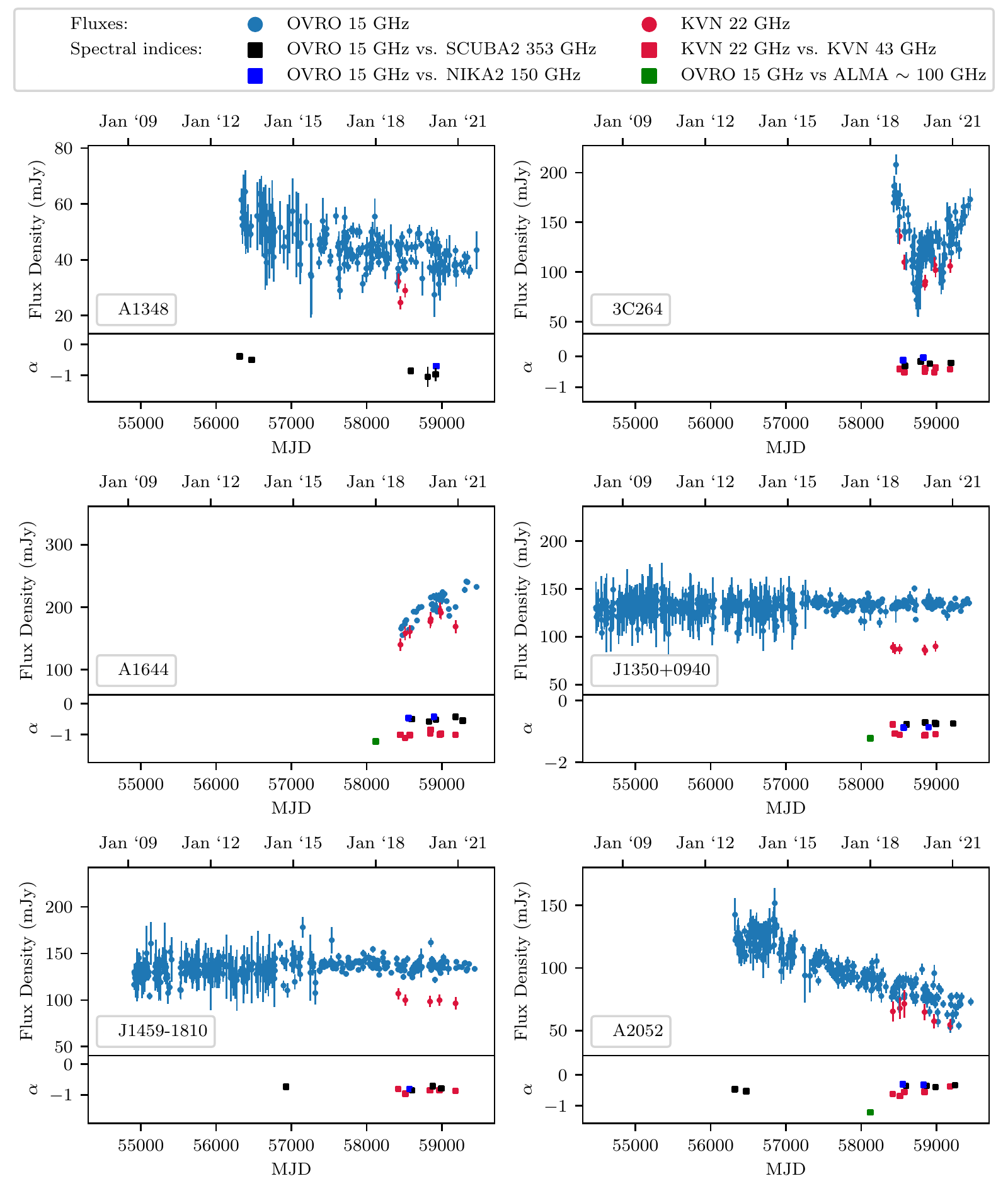}
     \caption[OVRO 15GHz lightcurves of BCGs]{Continued from Fig. \ref{fig:combined_lightcurves_1}. Continued in Figs. \ref{fig:combined_lightcurves_3} and \ref{fig:combined_lightcurves_4}.}
    \label{fig:combined_lightcurves_2}
\end{figure*}

\begin{figure*}
	\includegraphics[width=1\textwidth]{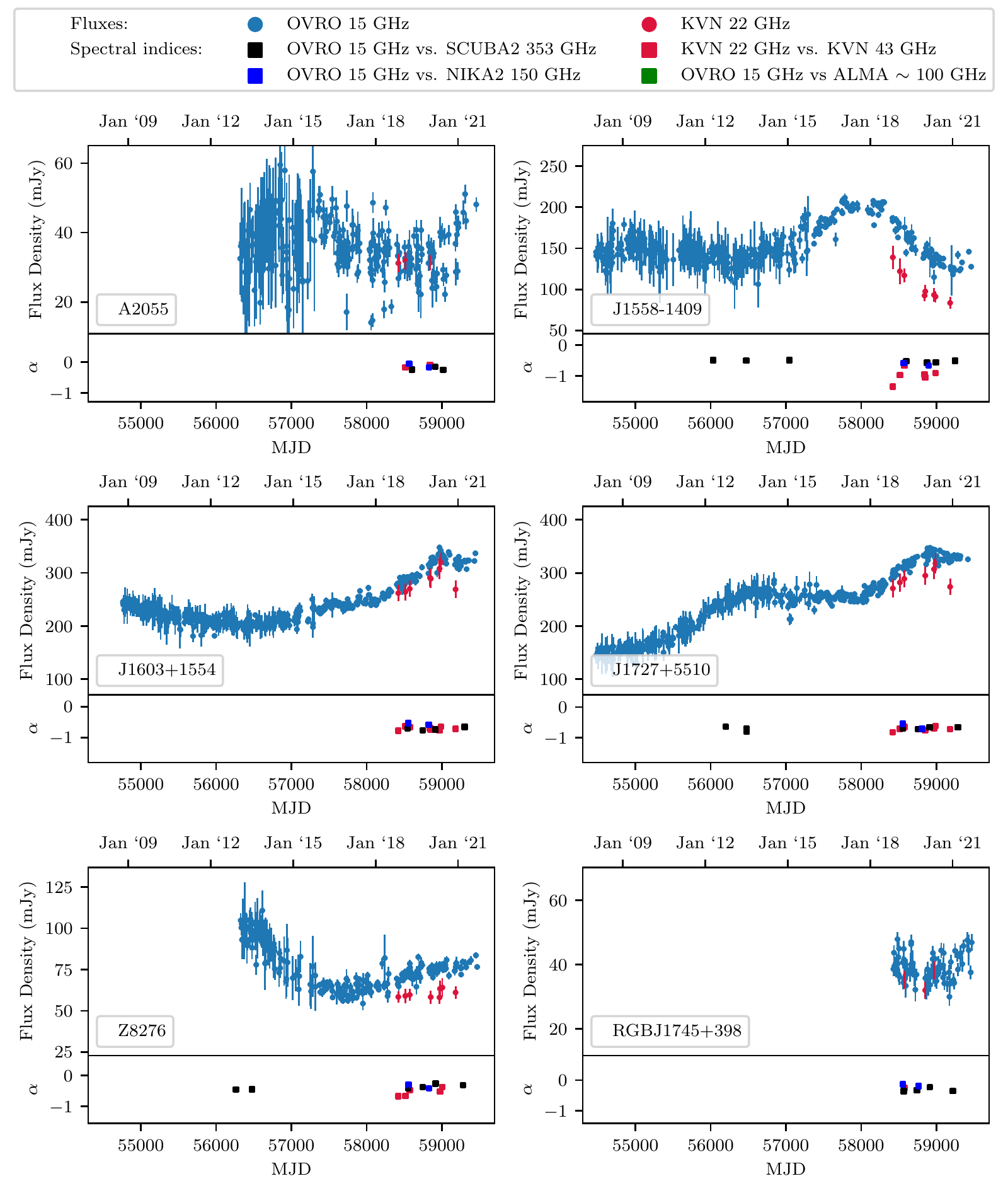}
     \caption[OVRO 15GHz lightcurves of BCGs]{Continued from Fig. \ref{fig:combined_lightcurves_1} and \ref{fig:combined_lightcurves_2}. Continued in Fig. \ref{fig:combined_lightcurves_4}.}
    \label{fig:combined_lightcurves_3}
\end{figure*}

\begin{figure*}
	\includegraphics[width=1\textwidth]{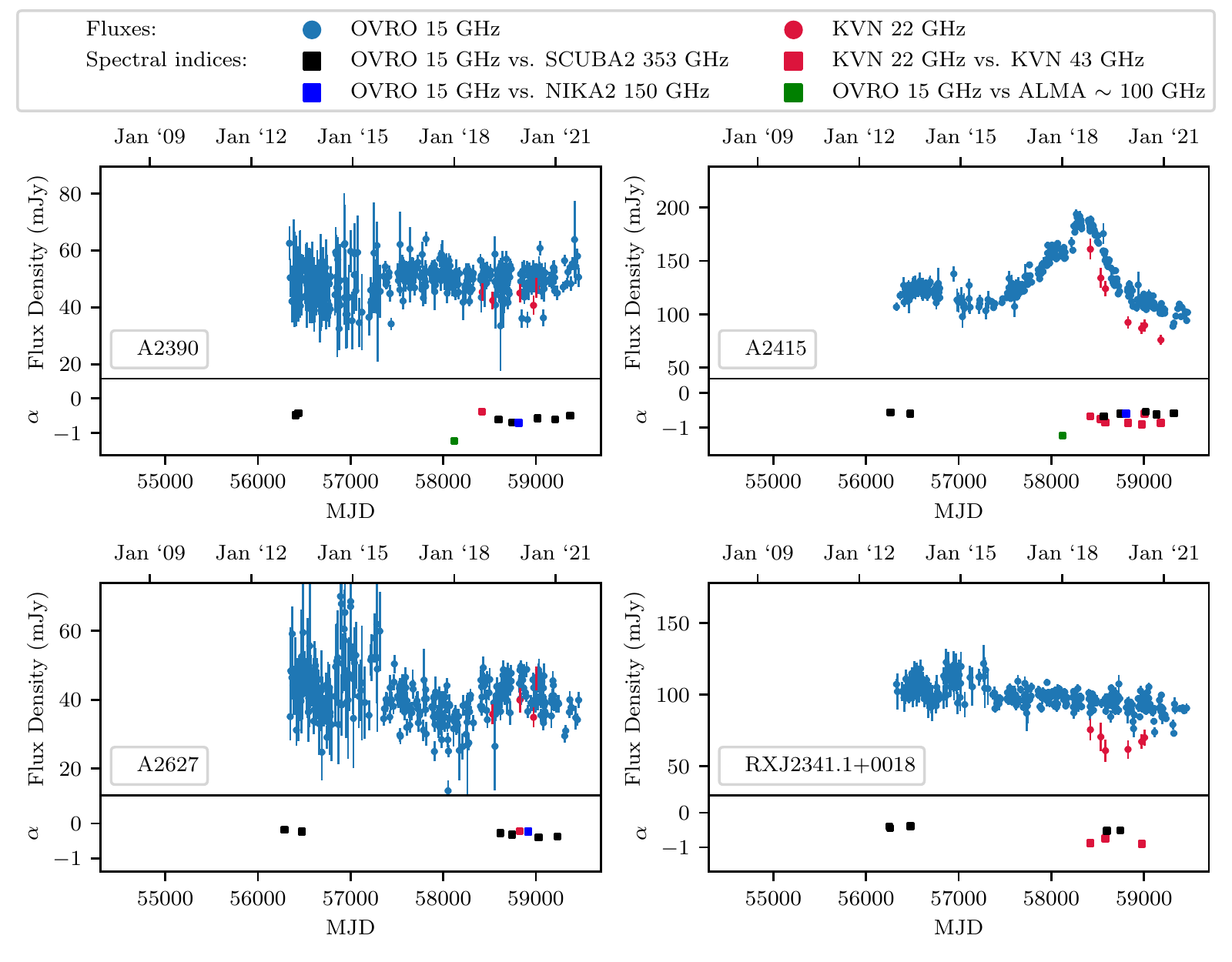}
     \caption[OVRO 15GHz lightcurves of BCGs]{Continued from Fig. \ref{fig:combined_lightcurves_1}, \ref{fig:combined_lightcurves_2}, and \ref{fig:combined_lightcurves_3}.}
    \label{fig:combined_lightcurves_4}
\end{figure*}

\subsection{Masking}
\label{sec:masking}

The OVRO lightcurves contain a small but significant number of spurious data points. These either have errors significantly larger than their neighbours, or are implausible measurements. To prevent this affecting the analysis, any observation, i, with flux density, $F_i$, and error, $\sigma_i$, which meets one or both of the following conditions is masked:
\begin{equation}
    \sigma_{i} > 3 \bar{\sigma}
\end{equation}

\begin{equation}
    \left| F_{i} - \frac{F_{i-2} + F_{i-1} + F_{i+1} + F_{i+2}}{4} \right| > 3\bar{\sigma}
\end{equation}
The first condition removes measurements with particularly large errors, while the second removes those which are significantly different to their neighbours. 
Re-inspection of the masked lightcurves, which are shown in Figs. \ref{fig:combined_lightcurves_1}, \ref{fig:combined_lightcurves_2}, \ref{fig:combined_lightcurves_3}, and \ref{fig:combined_lightcurves_4}, shows a large reduction in the number of spurious measurements (the unmasked spectra can be seen in Appendix \ref{sec:MaskingAppendix}). We find that this level of masking provides a good balance -- more conservative masking removes good data, while less strict masking leaves implausible measurements unmasked.

\subsection{2014 blip in lightcurves}

A spurious dimming and brightening appears in some of the lightcurves in 2014/15, most notably in the calibrator source DR21. This is visible in Fig. \ref{fig:combined_lightcurves_1}, and a similar feature can also be seen in J0439+0520. This effect coincides with the installation of a new receiver on the OVRO 40m telescope in May 2014.

The effect is not so apparent in any of the other sources, and is generally within the 5 per cent absolute calibration uncertainty. Nevertheless, to assess its impact, we applied the variability detection parameters to all the lightcurves with and without this period (mjd 56750 to 57400) being masked. In all cases the difference in the statistics is less than the rounding error, except for the calibrator source DR21. In the following analysis we therefore mask this region for DR21, but leave it unmasked for all other sources.

\subsection{Correction for extended emission}
\label{sec:extended_emission_correction}

\begin{table*}
\caption[KVN predictions of 15 GHz flux densities]{The 15 GHz flux densities predicted by 22 and 43 GHz KVN observations between the end of 2018 and the end of 2020, the mean value obtained with the single dish OVRO 40m telescope over the same period, and the resulting correction which has been applied to the spectra in Figs. \ref{fig:combined_lightcurves_1}, \ref{fig:combined_lightcurves_2}, \ref{fig:combined_lightcurves_3}, and \ref{fig:combined_lightcurves_4}.} 
    \label{tab:KVN_correction}
    \centering
    \begin{tabular}{lccccc}
\hline
Source & 15 GHz flux density & 15 GHz OVRO flux & Difference & Implied extended & Correction Applied\\
  & predicted by KVN / mJy & density / mJy & / mJy & emission / \% & / mJy \\
\hline
RXJ0132.6-0804  &  95  &  149  &  54  &  36 & 54 \\
J0439+0520  &  381  &  336  &  -45  &  -13 & 0 \\
A646  &  57  &  55  &  -2  &  -3 & 0 \\
4C55.16  &  514  &  1571  &  1057  &  67 & 1057 \\
A1348  &  --  &  40  &  --  &  -- & 0 \\
3C264  &  125  &  733  &  608  &  82 & 608 \\
A1644  &  250  &  200  &  -50  &  -25 & 0 \\
J1350+0940  &  130  &  134  &  4  &  2 & 0 \\
J1459-1810  &  140  &  136  &  -4  &  -2 & 0 \\
A2052  &  78  &  230  &  152  &  66 & 152 \\
A2055  &  33  &  116  &  83  &  71 & 83 \\
J1558-1409  &  148  &  204  &  56  &  27 & 56 \\
J1603+1554  &  373  &  306  &  -67  &  -21 & 0 \\
J1727+5510  &  381  &  323  &  -58  &  -17 & 0 \\
Z8276  &  73  &  74  &  1  &  1 & 0 \\
RGBJ1745+398  &  39  &  126  &  87  &  69 & 87 \\
A2390  &  50  &  89  &  39  &  43 & 39 \\
A2415  &  146  &  130  &  -16  &  -12 & 0 \\
A2627  &  42  &  99  &  57  &  57 & 57 \\
RXJ2341.1+0018  &  93  &  146  &  53  &  36 & 53 \\
\hline
    \end{tabular}
\end{table*}
The OVRO 40m Telescope with which the 15 GHz lightcurves were obtained is a single dish instrument, so its measurements are of low angular resolution. As a result, the flux densities may incorporate both the radio core and more extended radio lobe components. In cases where the radio lobes - which do not vary on human timescales - are particularly strong, they can dilute the apparent variability of the radio core.

Between around the end of 2018 and the end of 2020, each source was observed by KVN at 22 and 43 GHz. KVN is a very long baseline array, and is able to resolve out the extended emission. 

We use these higher resolution observations to correct for the extended emission in many of the OVRO 15 GHz flux densities. To do this, we use the mean 22 and 43 GHz flux densities to estimate the spectral index over the observed period. This spectral index is then used to predict the 15 GHz flux density. Table \ref{tab:KVN_correction} gives the 15 GHz flux densities predicted by KVN, and the values obtained over the same period by OVRO. The table also includes the difference between the two and the implied fraction of the emission which is extended. 

In many cases the predicted 15 GHz flux density is significantly lower than that found by OVRO due to the extended emission. Since we are only interested in the variability of the core component, we subtract this difference from every point on the 15 GHz OVRO lightcurves. However, we only do this for sources where the implied extended emission is more than 25 per cent of the total flux density measured by OVRO. This strikes a balance between removing significant amounts of extended emission, and over-correcting in cases with a non-uniform spectral index over the 15 to 43 GHz range. The correction applied to each lightcurve can be found in Table \ref{tab:KVN_correction}.

\section{Variability Detection and Quantification Parameters}
\label{sec:detection_techniques}

To characterize the variability in the 15 GHz lightcurves, we use a range of numerical parameters. 
Each has advantages and disadvantages, and several are required to give a reliable impression of a source's variability. We therefore use a combination of four to analyse the variability of the observed sources. Each of them is described in the following subsections. Many other methods for quantifying variability in photometric time series data exist, though they are often less effective when applied to AGN, or require many hundreds or thousands of observations to act as robust parameters \citep{Sokolovsky2017}. The four applied here are the ones we find to be most helpful when applied to our data.

\subsection{Reduced $\chi^{2}$ Test}

A $\chi^{2}$ test can assess the null hypothesis that a lightcurve is from a non-varying source, and therefore does not change in brightness. With a flat lightcurve model, the reduced $\chi^{2}$ of the null hypothesis is defined as 
\begin{equation}
    \chi^{2}_{r} = \frac{1}{N}\sum^{N}_{i=1} \left( \frac{F_{i} - \bar{F}}{\sigma_{i}}\right)^{2},
\end{equation}
where $N$ is the number of measurements of the source's flux density, $F_i$, and $\sigma_i$ is the measurement error. If the true lightcurve of the source is flat, this test gives the probability of observing a lightcurve with an apparent level of variability greater than or equal to that which is detected, assuming measurement errors are estimated correctly. 

\subsection{Variability Index}

The variability index, $VI$, is defined as
\begin{equation}
VI = \sqrt{\frac{\sum^{N}_{i=1}(F_{i} - \bar{F})^{2} - \sum^{N}_{i=1} \sigma_{i}^{2}}{N}},
\end{equation}
where $N$ is the number of measurements, $F_i$ is the flux density and $\sigma_i$ is the associated error. If the discriminant of the variability index is negative, the variability index is calculated with the modulus of the discriminant and given a negative sign. Such a result indicates that there is no variability between the two measurements. A variable source is expected to have a VI greater than zero, with larger positive values corresponding to stronger variability.

The most notable advantage of the variability index is that it can be used with as few as two flux density measurements. Since our OVRO observations have a median separation of seven days, this test can be used to assess variability on approximately week long timescales. However, as previously stated, variability cannot take place on timescales less than a source's light crossing time, so week-to-week variability is not expected. Nevertheless, we analyse the spectra on these timescales for completeness.

\subsection{Variability Amplitude}

The variability amplitude, $VA$, is used to quantify the magnitude of the peak-to-trough variability in a source's lightcurve. It is defined as \citep{Heidt1996}:
\begin{equation}
    VA (\%) = \frac{100}{\bar{F}} \sqrt{(F_{\textnormal{max}} - F_{\textnormal{min}})^{2} - 2\bar{\sigma^{2}}},
    \label{eq:va}
\end{equation}
where
\begin{equation}
    \bar{\sigma^{2}} = \frac{1}{N}\sum^{N}_{i=1}\bar{\sigma_{i}^{2}}.
\end{equation}
Here, $F$ represents the observed flux densities and $\sigma$ the associated errors. The variability amplitude indicates the size of the peak-to-trough variability, and is often calculated with only the single largest and smallest flux density measurements (i.e. $N=2$). 

However, where a light curve has a large number of data points, $F_{\textnormal{max}}$ and $F_{\textnormal{min}}$ can be replaced e.g. by the mean of the several largest and smallest flux densities. This makes it less prone to selecting outlying data points and overstating the variability in a lightcurve, particularly where the variability is small compared with the measurement error.

\subsection{Interquartile Range}

The interquartile range (IQR) of a lightcurve is found by splitting the data into two halves about the median, then taking the difference between the median value of the upper and lower halves. For a normal distribution, IQR = 1.35$\,\sigma$. The IQR is one of the most robust parameters for quantifying variability due to its low sensitivity to outlying data points. However, to be robust the IQR requires at least several tens of measurements, so with our 15~GHz data it is only useful for multi-year to decade long timescales \citep{Sokolovsky2017}.

\subsection{Mock Lightcurves}
\label{sec:mock_spectra}
The apparent variability in a source's lightcurve is a product of its true variability and the observational errors, which cannot be known precisely. For some of the weaker sources in our sample, the observational errors are clearly significant. For example, in A2055 of Fig \ref{fig:combined_lightcurves_3}, the single largest and smallest flux density measurements would suggest a peak-to-trough change in flux density of around 300 per cent in approximately 3 years, despite its lightcurve being relatively flat.

To estimate the true level of variability of each source, we create mock lightcurves for each source based on a flat lightcurve model. By applying the variability detection and quantification parameters described above to the real and mock lightcurves, we can more confidently determine the level of variability in each source.

For each of the 20 sources with OVRO 15 GHz observations, 1000 mock lightcurves are created. These can then be used to find the values given by the variability parameters for a truly flat lightcurve with similar flux density errors. By calculating how often the mock data result in variability greater than or equal to that of the true lightcurves, the probability that the observed variability has arisen as a result of only noise, rather than intrinsic source variability, can be found.

The mock lightcurves are created in the following way:
\begin{itemize}
    \item The real OVRO observations are split in two at \mbox{mjd = 56800}, the approximate date at which the new phase calibrator was introduced and there was a large reduction in the observational errors. This is done after masking bad data and the removal of extended emission.
    \item The mean flux density, $\bar{F}$, and flux density error, $\bar{\sigma_{F}}$, of these two parts of the data are calculated.
    \item Mock flux densities are drawn for the two parts from a Gaussian distribution, centred on $\bar{F}$ and with a standard deviation of $\bar{\sigma_{F}}$. Each mock flux density is given an error of $\bar{\sigma_{F}}$. One mock flux density is drawn for each date at which there was a real observation.
\end{itemize}

In Appendix \ref{sec:MockSpectraAppendix}, we show one mock spectrum for each source for illustrative purposes.

\section{Application of Variability Detection and Quantification}
\label{sec:application_of_variaibility_detection_parameters}
The numerical parameters described in the previous section each have their own strengths and weaknesses, and all have a particular use when applied to the data. As such, we use:
\begin{itemize}
\item the reduced $\chi^{2}$ test to determine whether or not each lightcurve is consistent with that of a non-varying source
\item the variability index, VI, to detect variability on short to medium term timescales (weeks to years)
\item the variability amplitude, VA, to show the peak-to-trough changes in the lightcurves over medium to long term timescales (year to multi-year)
\item the interquartile range, IQR, to categorize each source's variability as undetectable, weak, moderate, strong, or very strong over the full timespan on which it has been observed.
\end{itemize}

Results from the application of the variability detection parameters to the real and mock lightcurves are described in the following subsections.

\subsection{Reduced $\chi^{2}$ Test -- determining consistency with a flat lightcurve model}
The reduced $\chi^{2}$ and a corresponding p-value are calculated using the full extent of each source's lightcurve (after masking and the removal of extended emission), and the resulting values are shown in Table \ref{tab:variability_tests}. This p-value does not give the probability that a source is non-varying over the observed period. Instead, it gives the probability of obtaining a lightcurve at least this discrepant from the flat lightcurve model, assuming the model is true and the errors are accurately estimated. 

If the p-value is $<5$ per cent, we consider the lightcurve to be inconsistent with the flat lightcurve model. According to this criterion, 15 of the sources are inconsistent with a flat lightcurve model, while 5 are consistent.

\begin{table*}
	\caption[Results of variability tests on 15 GHz lightcurves]{Results from the four variability detection techniques applied to the 15 GHz OVRO lightcurves. Bold typeface is used for values indicating variability. \\ 1(a) The p-value given by the $\chi^{2}$ test using a flat lightcurve model. \\ 2(a-d) The percentage of mock lightcurves for which the median VI at 10, 30, 100, and 300 days exceeds that of the real spectra. \\ 3(a-d) The percentage of mock lightcurves for which the median VA at 100, 300, 1000, and 3000 days exceeds that of the real spectra. \\ 4(a) The percentage of mock lightcurves for which the IQR (calculated over the full timespan of the lightcurve) exceeds that of the real spectra. \\ 4(b) The IQR of each real lightcurve divided by the mean flux density \\ 4(c) The variability classification of each source. This is determined by comparing the IQR/${\bar{F}}$ value to that of the calibrator sources (see \S\ref{sec:application_of_variaibility_detection_parameters}).}
	\begin{tabular}{lcccccccccccc} 
\hline
 & (1) $\chi^{2}$ test & \multicolumn{4}{c}{(2) VI} & \multicolumn{3}{c}{(3) VA} & \multicolumn{3}{c}{(4) IQR} \\
Source & (a) p-value /$\%$  &  (a) 10   &  (b) 30   &  (c) 100  & (d) 300  &  (a) 300   &  (b) 1000  & (c) 3000 & (a) t$_{\textnormal{max}}$ & (b) $/ \bar{F}$ & (c) Classification\\
\hline
3C286 - Calibrator & 11.1 & 67.4 & 58.7 & 29.9 & \textbf{1.8} & \textbf{0.3} & \textbf{<0.1} & \textbf{<0.1} & \textbf{<0.1} & 0.02 & negligible \\
DR21 - Calibrator & 12.9 & 50.5 & 22.5 & \textbf{3.1} & \textbf{<0.1} & \textbf{0.2} & \textbf{<0.1} & \textbf{<0.1} & \textbf{<0.1} & 0.02 & negligible \\
RXJ0132.6-0804 & \textbf{1.1} & 22.9 & 33.8 & \textbf{4.6} & \textbf{<0.1} & \textbf{<0.1} & \textbf{<0.1} & \textbf{<0.1} & \textbf{<0.1}& 0.13 & \textbf{moderate} \\
J0439+0520 & \textbf{0.6} & 46.4 & 74.4 & 47.7 & \textbf{0.3} & \textbf{1.4} & \textbf{<0.1} & \textbf{<0.1} & \textbf{<0.1} & 0.08 & \textbf{weak} \\
A646 & \textbf{0.1} & 30.7 & 50.4 & \textbf{4.2} & \textbf{<0.1} & \textbf{0.2} & \textbf{<0.1} & \textbf{<0.1} & \textbf{<0.1} & 0.17 & \textbf{strong} \\
4C55.16 & \textbf{2.5} & 42.3 & 69.6 & 33.6 & \textbf{0.2} & \textbf{1.8} & \textbf{<0.1} & \textbf{<0.1} & \textbf{<0.1} & 0.1 & \textbf{weak} \\
A1348 & \textbf{4.8} & 19.5 & 50.4 & 66.5 & 44.7 & 38.5 & \textbf{0.15} & \textbf{<0.1} & \textbf{<0.1} & 0.22 & \textbf{strong} \\
3C264 & \textbf{0.8} & 22.4 & 27.2 & 5.9 & \textbf{<0.1} & \textbf{<0.1} & \textbf{<0.1} & \textbf{<0.1} & \textbf{<0.1} & 0.24 & \textbf{strong} \\
A1644 & \textbf{<0.1} & 6.0 & 8.8 & \textbf{<0.1} & \textbf{<0.1} & \textbf{<0.1} & \textbf{<0.1} & \textbf{<0.1} & \textbf{<0.1} & 0.12 & \textbf{moderate} \\
J1350+0940 & 24.7 & 57.0 & 79.7 & 79.2 & 94.2 & 99.6 & 100.0 & 87.9 & 16.0 & 0.06 & \textbf{weak} \\
J1459-1810 & 13.1 & 95.3 & 71.7 & 44.8 & 9.3 & 33.2 & \textbf{1.6} & \textbf{<0.1} & \textbf{<0.1} & 0.07 & \textbf{weak} \\
A2052 & \textbf{<0.1} & 23.0 & 61.5 & 64.9 & 81.0 & 64.3 & \textbf{<0.1} & \textbf{<0.1} & \textbf{<0.1} & 0.3 & \textbf{very strong} \\
A2055 & \textbf{3.0} & 16.2 & 27.4 & 13.8 & 6.8 & 10.2 & \textbf{<0.1} & \textbf{<0.1} & \textbf{<0.1} & 0.26 & \textbf{very strong} \\
J1558-1409 & \textbf{<0.1} & 41.7 & 63.3 & 60.4 & 72.8 & 64.2 & 99.2 & \textbf{<0.1} & \textbf{<0.1} & 0.14 & \textbf{moderate} \\
J1603+1554 & \textbf{<0.1} & 59.1 & 85.6 & 98.0 & 84.9 & 100.0 & \textbf{<0.1} & \textbf{<0.1} & \textbf{<0.1} & 0.15 & \textbf{strong} \\
J1727+5510 & \textbf{<0.1} & 100.0 & 79.5 & 49.8 & \textbf{0.7} & 73.5 & \textbf{<0.1} & \textbf{<0.1} & \textbf{<0.1} & 0.42 & \textbf{very strong} \\
Z8276 & \textbf{<0.1} & 23.4 & 49.9 & 16.1 & \textbf{0.6} & \textbf{0.73} & \textbf{<0.1} & \textbf{<0.1} & \textbf{<0.1} & 0.26 & \textbf{very strong} \\
RGBJ1745+398 & 9.5 & 18.7 & 15.3 & \textbf{0.5} & \textbf{<0.1} & \textbf{<0.1} & \textbf{<0.1} & \textbf{<0.1} & \textbf{<0.1} & 0.14 & \textbf{moderate} \\
A2390 & 16.1 & 32.2 & 63.0 & 30.4 & \textbf{1.6} & \textbf{<0.1} & \textbf{<0.1} & \textbf{<0.1} & \textbf{<0.1} & 0.11 & \textbf{moderate} \\
A2415 & \textbf{<0.1} & 35.2 & 52.2 & 42.6 & 18.3 & 30.2 & \textbf{<0.1} & \textbf{<0.1} & \textbf{<0.1} & 0.24 & \textbf{strong} \\
A2627 & 5.0 & 32.5 & 37.0 & 6.7 & \textbf{<0.1} & \textbf{<0.1} & \textbf{<0.1} & \textbf{<0.1} & \textbf{<0.1} & 0.19 & \textbf{strong} \\
RXJ2341.1+0018 & \textbf{4.6} & 25.7 & 53.9 & 49.5 & \textbf{<0.1} & \textbf{0.14} & \textbf{<0.1} & \textbf{<0.1} & \textbf{<0.1} & 0.09 & \textbf{weak} \\
\hline

	\end{tabular}
	\label{tab:variability_tests}
\end{table*}

\subsection{Variability Index -- detecting variability on short timescales}

The VI is useful in cases where there are a small number of data points, and can be used even with just two flux density measurements. We use it to indicate whether each source shows variability on approximately week to year long timescales. To do this, the parameter is calculated at every available 10, 30, 100, and 300 day timespan within the observations. For example, for the 10 day timespan, the VI is calculated with measurements from between days 1 and 10, 2 and 11, 3 and 12 etc. of the observations, assuming there is a unique set of at least two observations within those time periods.

The percentage of the 1000 mock lightcurves for which the median VI at 10, 30, 100, and 300 days exceeds that of the real lightcurves is then used as a threshold to determine whether or not variability is present. If 5 per cent or fewer of the mock lightcurves result in a VI larger than that of the real lightcurves, the source is classed as variable. 

The percentage of the mock lightcurves for which the VI at 10, 30, 100, and 300 days exceeds that of the real lightcurves is shown in Table \ref{tab:variability_tests}. These VI tests suggest that no sources vary on 10 or 30 day timescales, four vary on 100 day timescales, and 13/20 sources vary on 300 day timescales.

\subsection{Variability Amplitude -- peak-to-trough variability}

We use the VA to determine the magnitude of the peak-to-trough changes in the lightcurves over timescales of 100 to 3000 days. Due to the large number of observations, in our calculations we follow the example of \citet{Hogan2015b} and use the mean of the five largest and smallest flux densities in place of $F_{\textnormal{max}}$ and $F_{\textnormal{min}}$, rather than the single largest and smallest values. This approach provides a good compromise between reducing the effect of outlying flux densities, and the inclusion of more flux densities with less extreme separation reducing the apparent VA. 

The percentage of the mock lightcurves for which the VA at 300, 1000, and 3000 days exceeds that of the real lightcurves is shown in Table \ref{tab:variability_tests}. Excluding calibrator sources, and with a threshold of 5 percent, this suggests that 11/20 sources typically vary on 300 day timescales, 18/20 vary on 1000 day timescales, and 19/20 typically vary on 3000 day timescales\footnote{Three sources (3C264, A1644, and RGBJ1745+398) only have observations spanning 2.8 years, so the VA values calculated at 1000 and 3000 days are lower limits. However, these still meet the variability threshold, in spite of the shorter timespan of the observations.}. 

Figs. \ref{fig:variability_amplitudes_1} and \ref{fig:variability_amplitudes_2} also show the VA for timespans between approximately 150 and 3600 days for both the real spectra, and the mock lightcurves. As was the case for the variability index, the parameter is calculated at every available timespan within the observations. For example, for the 150 day timespan, the VA is calculated between days 1 and 150, days 2 and 151, days 3 and 152 etc.. 

\begin{figure*}
	\includegraphics[width=\textwidth]{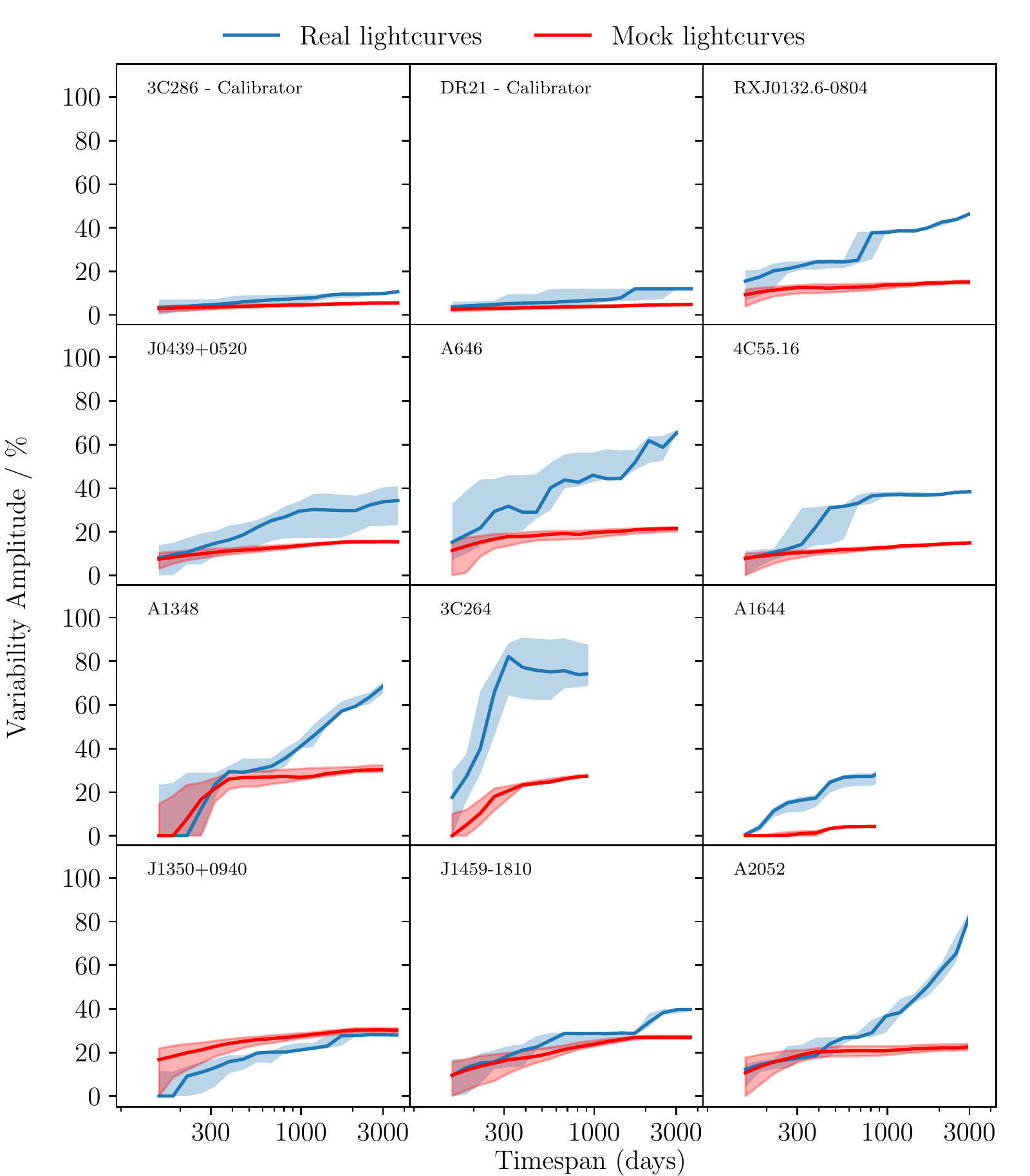}
     \caption[Variability amplitudes from 150 to 3600 days]{Variability amplitudes for timespans of 150 to 3600 days. In most cases, the total time range over which the OVRO 15 GHz observations were carried out is several years, and the dark blue line represents the median VA within that time period. The shaded blue regions mark the maximum and minimum VA within the full period over which each source was observed. The red lines show similar results, but for the mock flat lightcurves. Note that 3C264 and A1644 only have observations spanning 2.8 years ($\sim 1000$ days).}
    \label{fig:variability_amplitudes_1}
\end{figure*} 

\begin{figure*}
	\includegraphics[width=\textwidth]{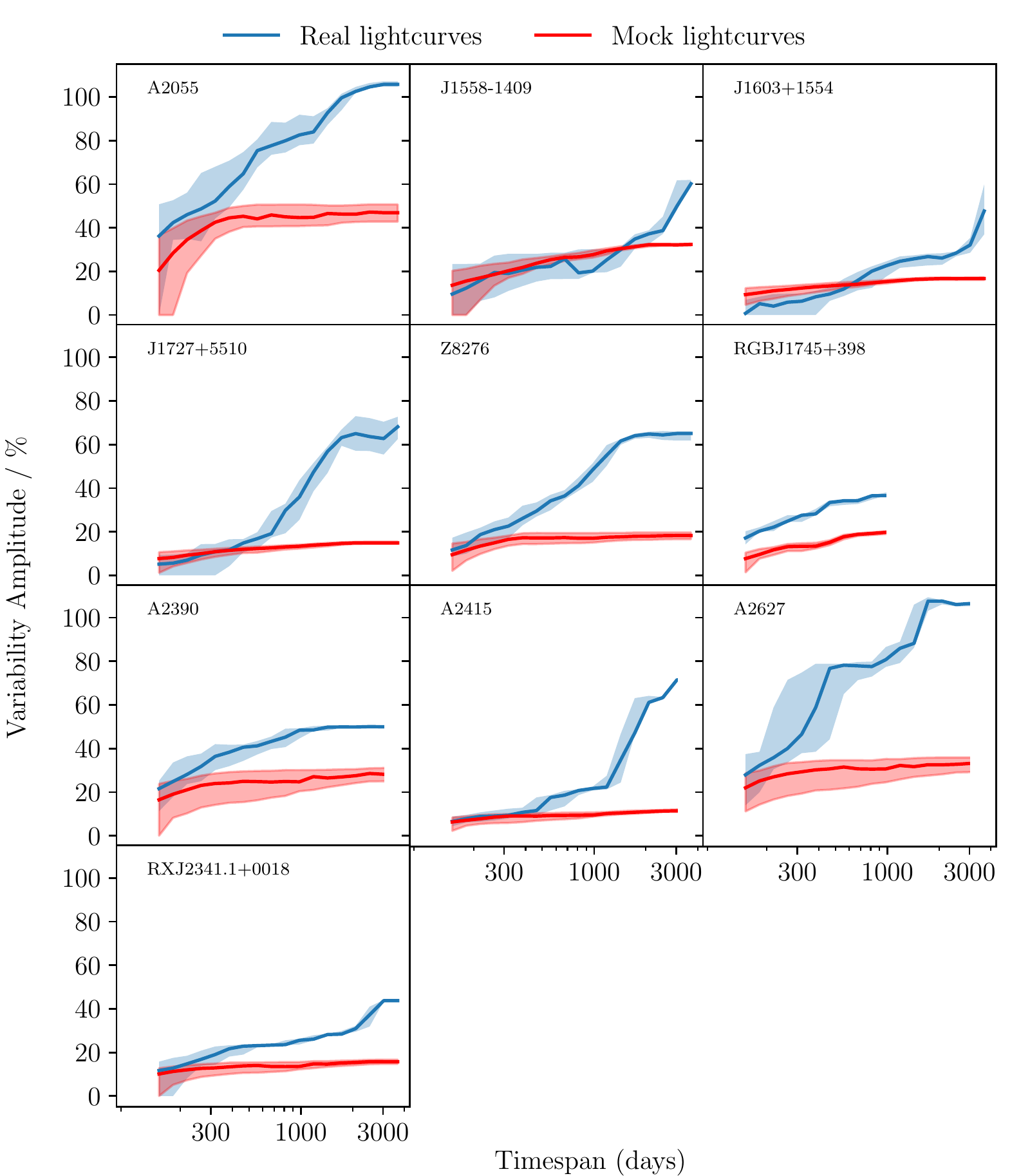}
     \caption[Variability amplitudes from 150 to 3600 days]{Continued from Fig. \ref{fig:variability_amplitudes_1}. Note that RGBJ1745+398 only has observations spanning 2.8 years ($\sim 1000$ days).}
    \label{fig:variability_amplitudes_2}
\end{figure*}

The dark blue lines in Figs. \ref{fig:variability_amplitudes_1} and \ref{fig:variability_amplitudes_2} represent the median VA of the real spectra, while the edges of the shaded blue regions mark the maximum and minimum VA seen within the observations. Similarly, the red lines represent the mock lightcurves. These red lines show that observational errors are responsible for a significant fraction of the apparent VA in many sources, particularly on shorter timespans. However, the excess VA (i.e. the gap between the red and blue lines) is also a clear indicator of variability in many sources. 

The width of the shaded regions also indicates how consistent a source's variability is. For example, the range in the VA values calculated for the calibrator source 3C286, which has a consistently flat lightcurve, is very narrow. Similarly, the range of VAs calculated for A2052 is narrow across all timespans because the source varies at a close to constant rate throughout the whole period of observations. On the other hand, the lightcurve of e.g. A646 contains some periods which are much less variable than others, resulting in the wide range of VAs shown by the blue shaded region in Fig. \ref{fig:variability_amplitudes_1}.

\subsection{Interquartile Range -- quantifying variability over the longest available timescales}

The IQR is perhaps the most robust parameter for detecting variability, but is only reliable with several tens of flux measurements \citep{Sokolovsky2017}. Over the full period in which each source was observed, we therefore use the IQR to categorize each source as having variability which is negligible, weak, moderate, strong, or very strong. In most cases the observations span several years or more, and in this way we are able to classify the strength of the variability seen in each source based on the entirety of the available data.

Where the value of $\textnormal{IQR}/\bar{\textnormal{F}}$ is less than three times the value of the calibrator sources, i.e. $<0.06$, the variability is described as negligible. Otherwise, if $\textnormal{IQR}/\bar{\textnormal{F}}$ is in the range $0.06 - 0.10$ (i.e. three to six times that of the calibrator sources), it is labelled as weak, $0.10 - 0.15$ as moderate, $0.15 - 0.25$ as strong, and $>0.25$ as very strong. The IQR and variability classification for each source's lightcurve are shown in Table \ref{tab:variability_tests} (as well as the percentage of the mock spectra for which the IQR exceeds that of the real lightcurves).

According to this variability characterization parameter, we find that no source has negligible levels of variability over the full observation timespan, while 5/20 sources show weak levels of variability, 5/20 show moderate levels of variability, 6/20 show strong levels of variability, and 4/20 show very strong levels of variability.

\begin{figure*}
	\includegraphics[width=\textwidth]{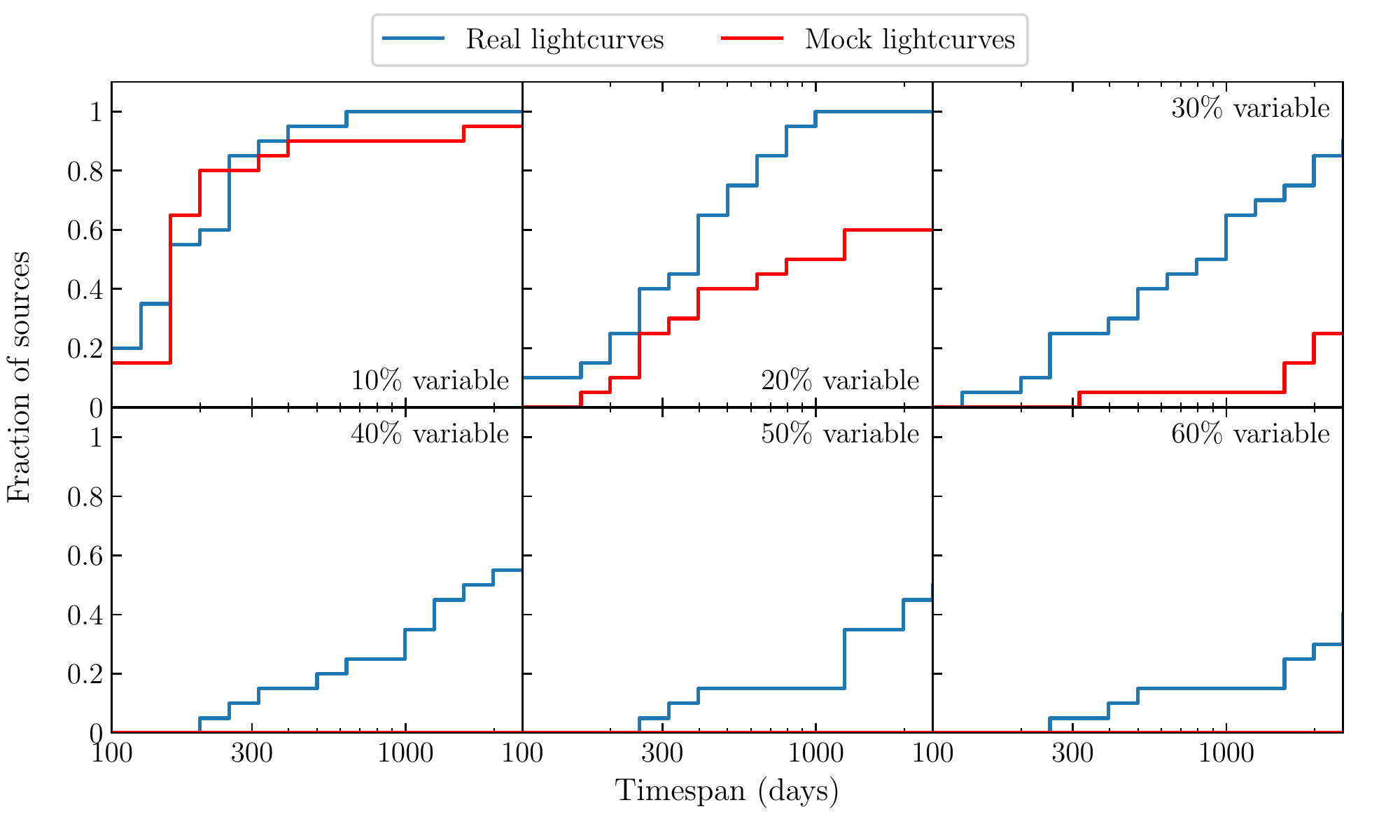}
     \caption[Fraction of sources varying by 10 to 60 per cent]{The fraction of sources for which the median peak-to-trough variability exceeds 10, 20, 30, 40, 50, and 60 per cent, calculated as a function of the observation timespan. The peak-to-trough variability is determined by the VA, given in equation \ref{eq:va}. Calibrator sources are not included. Note that 3C264, A1644, and RGBJ1745+398 only have observations spanning 2.8 years ($\sim 1000$ days).}
    \label{fig:fraction_of_sources_displaying_variaibility}
\end{figure*}

\section{Discussion}
\label{sec:discussion}
\subsection{Overall variability and implications for the Sunyaev-Zel'dovich Effect}

The precursor to this work, \citet{Hogan2015}, studied the high radio frequency variability of 16 brightest cluster galaxies on timescales of between 1.5 and 6.5 years. The size and high temporal sampling presented here is a significant improvement on this, so it is now possible to draw wider conclusions about the magnitude and ubiquity of variability in brightest cluster galaxy cores.

Fig. \ref{fig:fraction_of_sources_displaying_variaibility} shows the fraction of 
sources which vary by between 10 and 60 per cent over a range of timespans from 100 to 2500~days.
All sources appear to show 10 per cent peak-to-trough variability on 18~months timescales. However, due to noise, the same is true of the mock lightcurves and so at this level any true variability cannot be separated from the apparent variability caused by observational errors. 

For stronger levels of variability, it is easier to disentangle the true variability from the noise of the spectra. Peak-to-trough changes of 20 per cent are seen in all sources on 3~year timescales versus half of the mock lightcurves, so moderate levels of variability are a feature of at least half these sources. Similarly, at least a third of the sources vary at the 60 per cent level on 6~year timescales. This is important because studies of Sunyaev-Zel'dovich Effect in the radio/sub-mm often involve the subtraction of the radio/sub-mm continuum source from the BCG. This subtraction is often made by extrapolating from a lower frequency radio flux density made at least a few years before the Sunyaev-Zel’dovich observations. The complex spectral nature of these radio sources and their intrinsic variability means that these estimates of flux density will be far less certain than assumed. 

The OVRO lightcurves themselves show significant changes in variability over the 3--12~years covered. For instance A2415 and J1558-1409 show 3--4~year "outbursts" and several other sources transition from declining to rising over the full lightcurve.

In addition to the sources with OVRO 40m monitoring, we also have multiple high frequency observations of a number of additional sources that were either too far south to be observed with the OVRO 40m telescope (dec$<-20^\circ$) or too faint ($<100$~mJy at 15~GHz), but significant variability is implied. This is most notable are (i) A3581, the brightest source detected with SCUBA2, which shows an 80 per cent increase in flux density at 363~GHz between 2013 and 2021, and (ii) Zw8193, which has brightened by a factor of 2 at 150~GHz between 2011 and 2020. The seven other sources with multiple observations do not vary by more than 30 per cent at any frequency. This includes RXJ1745+39, E1821+644 and RXJ1832+68, which have limited Mets\"ahovi 37~GHz monitoring between 2002 and the present day, but the upper limits obtained are consistent with no significantly brighter emission in the recent past for these sources (L\"ahteenm\"aki, priv comm). This proportion is consistent with the fraction of sources (2 of 9) expected from Fig. \ref{fig:fraction_of_sources_displaying_variaibility} for the largest amplitude of variability.

\subsection{Variability and Spectral Index}

The only radio core in a BCG which has a lightcurve with better long term sampling than the OVRO BCG sample presented here is NGC1275/3C264 in the Perseus cluster. This source has varied by over a factor of ten at 30--90~GHz over the 50 years since its first detection at these frequencies \citep{Dutson2013}. The overall variation of peaks in activity with a 30--40~year spacing, short term "flares" during these peaks and then long periods of decline after a peak are consistent with the variety, timescale and amplitude of the variability seen in this sample. Importantly, \citet{Dutson2013} show that the high frequency spectral index of the core changes significantly over this cycle of activity. As would be expected, the spectral index above 30~GHz is relatively flat when the activity is high and the source is increasing in brightness, then is steeper as it fades. 

\begin{figure}
	\includegraphics[width=\columnwidth]{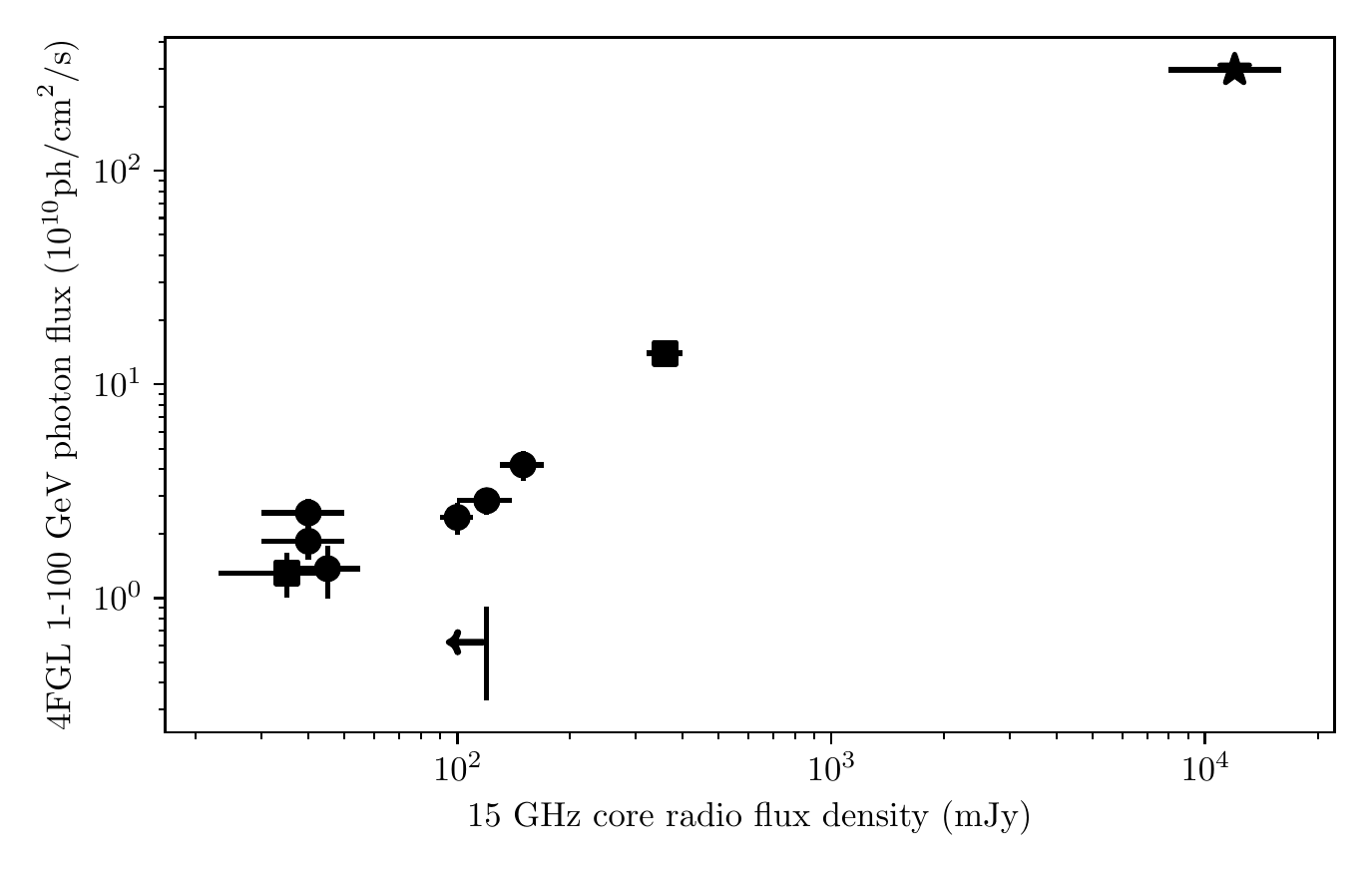}
     \caption[Fermi photon flux vs. 15 GHz core flux]{The {Fermi} 4FGL 1--100~GeV photon flux plotted against the mean 15~GHz core flux density for the six OVRO monitored sources (circles), NGC1275 (star) and three southern BCGs (squares and upper limit). Note that there is variability in the properties in both axes, but there is consistency between the two flux densities over the wide range.}
    \label{fig:4FGL}
\end{figure}

This dramatic change in the high frequency spectral index also mirrors the gamma-ray emission in NGC1275. \cite{Dutson2013} note the lack of observable gamma-ray emission from NGC1275 in the 1990s with EGRET on \textit{Compton} when the source was in strong decline but very clearly detected in the 1980s with \textit{COS-B} and after 2008 with \textit{Fermi}. Six of our OVRO sample have a significant \textit{Fermi} 4FGL detection in the first 8 years of data: 3C264/NGC3862, RXJ0132-08, A2055, RXJ1558-14, RXJ1745+39, and A2627 \citep{Abdollahi2020}. There are also three other BCGs in our parent X-ray sample with a 4FGL counterpart: A3112, A3392, and A3880. Using our mean OVRO 15~GHz flux density and literature values for the three additional sources and NGC1275 we can plot the core radio flux density at 15~GHz against the 4FGL photon flux (Fig. \ref{fig:4FGL}). There are few data points, but there is a clear correlation between these flux densities indicating that these other BCGs show radio to gamma-ray emission scaling as NGC1275.

\begin{figure}
	\includegraphics[width=\columnwidth]{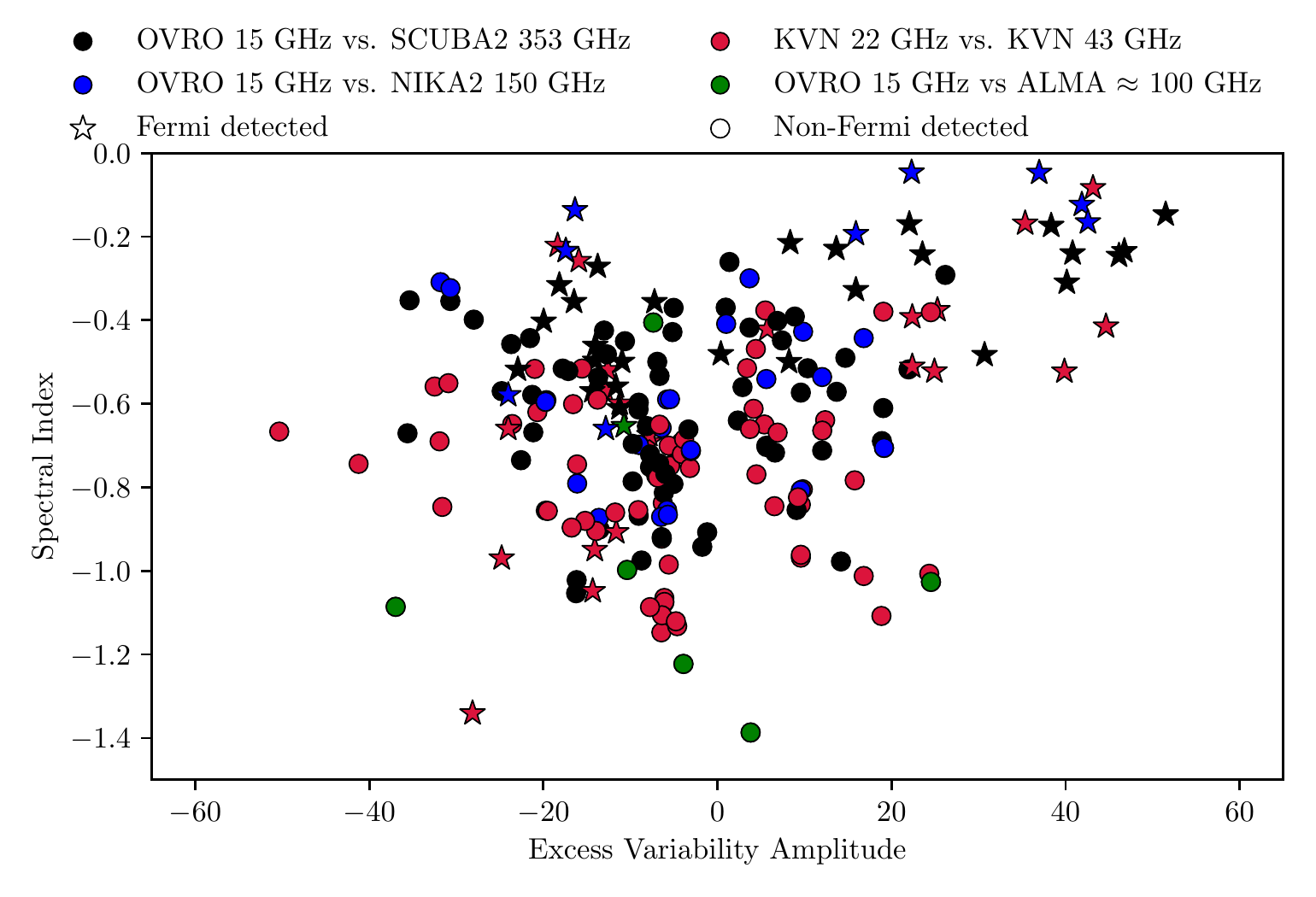}
     \caption[Spectral index vs. variability amplitude for BCGs]{The relationship between the spectral index and excess variability amplitude (i.e. the median VA of the real lightcurve minus the median VA of the real lightcurves) in the following two years. A negative excess VA indicates a fading lightcurve.}
    \label{fig:spectralindex_vs_variability_amplitude}
\end{figure}

With this link to the gamma-ray and the long timescale behaviour of NGC1275 in mind, we now consider the connection between variability and spectral index for our sample in more detail.
Fig. \ref{fig:spectralindex_vs_variability_amplitude} shows the spectral index between 15~GHz and our higher frequency KVN, SCUBA2, NIKA2, and ALMA photometry plotted against the excess amount that the source varies over the mock lightcurves in the subsequent two years (i.e. the median VA of the real lightcurve minus the median VA of the mock lightcurves). While the correlation is only moderate to low -- with a Pearson’s correlation coefficient of $0.30\pm0.02$ -- there is an overall trend that the spectral indices are flatter as sources increase in brightness. A strict correlation between the variability and spectral index is not necessarily expected due to the intricacies of AGN accretion and feedback mechanisms. For example, the link may be somewhat broken by the complex nature of the formation, expulsion, and breakdown of radio jets, and synchotron opacity effects within the core.

The sources with a {\it Fermi} 4FGL detection are plotted as stars and these are almost exclusively found in the upper part of the plot with a flatter spectral index but not always brightening. This is consistent with the behaviour of NGC1275 where the gamma-ray emission persists over the shorter term variability on year long timescales and only during the long decline, when the high frequency spectral index steepens, does the gamma-ray emission drop appreciably. The general lack of {\it Fermi} detections of sources with steep high frequency spectral indices -- despite having a 15~GHz flux density comparable to the sources with a 4FGL detection -- suggests that these are sources in the low activity part of a cycle similar to that seen in NGC1275. 

\citet{Hogan2015b} use archival data for two sources, 4C+55.16 and RXCJ1558-14, to show that at least some of our sample have been brighter in the period since the start of radio astronomy. However, sources that have brightened and should enter the sample will continue to be overlooked until we enter the era of regular, high frequency radio surveys \citep[the notable exception is the AT20G survey that selected all the sources in the equatorial overlap area between ATCA and OVRO between declinations of -20 and +0$^\circ$][]{murphy2010}.

The variation seen in NGC1275 on longer timescales than sampled here suggests that some of the sources in this sample will fade below the OVRO detection limit and other sources will brighten sufficiently to be monitored over the next 3--5~years \citep[][]{Dutson2014}. This makes it worth noting that because of the amplitude of the variability seen, a complete sample of sources for a study like this would be impossible since there will be sources continually rising above and fading below the radio flux density limit. However, we believe that our selection is representative and the properties of these sources can be used to constrain the duty cycle of activity in BCGs in general and act as a comparison for gamma-ray detected radio galaxies more widely.


\section{Conclusions}
\label{sec:conclusions}
The high radio frequency variability of brightest cluster galaxies provides important clues about the nature of fuelling and feedback cycles in galaxy clusters, but this behaviour is not well understood due to a lack of observational data.

To address this, we have carried out the most comprehensive high frequency monitoring of brightest cluster galaxies to date. This has primarily been done using the OVRO 40m telescope at 15 GHz, with additional observations from KVN at 22 and 43 GHz, NIKA2 at 150 GHz, SCUBA2 at 353 GHz, and ALMA at $\sim 100$ GHz. 

A range of variability detection and quantification parameters show significant levels of variability in most sources. Over the full timespan on which observations have so far been carried out, typically 8-13 years, 15/20 are inconsistent with a flat lightcurve model. Variability index and variability amplitude tests show that no sources vary on timescales of 10 or 30 days, while 4/20 vary on timescales of 100 days. At 300 days, 12-13/20 sources show evidence of variability, and at 18-19/20 sources vary on timescales of 1000 days. 

At least a third of our sample display 60 per cent variability on 6 year timescales. This is important to studies of the Sunyaev-Zel’dovich effect in the radio/sub-mm, which frequently involve the subtraction of the radio/sub-mm continuum source from the BCG. This is often done by extrapolating from a lower frequency radio flux density made several years before the Sunyaev-Zel’dovich observations. Due to the complex spectral nature of these radio sources and their intrinsic variability, estimates of flux density at higher frequencies will be far less certain than assumed.

We find a weak link between variability and spectral index changes in our sample, with the two being positively correlated. This is particularly true for sources with the strongest levels of variability, where spectral indices tend to be much flatter. This weak trend is similar to that seen in NGC 1275, where fading periods are associated with steep spectral indices, and outbursts with flatter spectral indices \citep{Dutson2013}.

Many of the sources have gone through phases in which they have varied significantly on periods of around three years, before returning to a relatively constant state (e.g. J0439+0520, J1558-1409, and J1727+5510). This behaviour confirms the compact nature of the continuum sources, and can be used as an upper limit on their physical size because variability cannot take place on timescales shorter than the light-crossing time. Simply observing the minimum timescale on which each source typically displays variability (see Table \ref{tab:variability_tests}) gives an upper limit of 1 - 3 ly (0.3-1 pc) in most cases. The `on-off' nature of the variability is also consistent with models of clumpy accretion \citep[e.g.][]{Pizzolato2005, Gaspari2018} and with observations of parsec scale clouds in the cores of massive galaxies like those studied here \citep[][]{David2014, Tremblay2016,Rose2019a, Rose2019b, Ruffa2019, Rose2020}.

\section*{Acknowledgements}
We thank the anonymous reviewer for their time and comments, which have helped to improve the paper.

This research has made use of data from the OVRO 40m monitoring program which was supported in part by NASA grants NNX08AW31G, NNX11A043G and NNX14AQ89G, and NSF grants AST-0808050 and AST-1109911, and private funding from Caltech and the MPIfR.

We are grateful to the staff of the KVN who helped to operate the array and to correlate the data. The KVN and a high-performance computing cluster are facilities operated by the KASI (Korea Astronomy and Space Science Institute). The KVN observations and correlations are supported through the high-speed network connections among the KVN sites provided by the KREONET (Korea Research Environment Open NETwork), which is managed and operated by the KISTI (Korea Institute of Science and Technology Information). The following people contributed to the acquisition and analysis of the KVN data: Jae-Woo Kim and Tae-Hyun Jung. Their affiliations are given on the title page of this paper.

We would like to thank the IRAM staff for their support during the numerous campaigns. The NIKA2 dilution cryostat has been designed and built at the Institut Néel.

We are grateful to the staff of the James Clerk Maxwell Telescope, which is operated by the East Asian Observatory on behalf of The National Astronomical Observatory of Japan; Academia Sinica Institute of Astronomy and Astrophysics; the Korea Astronomy and Space Science Institute; Center for Astronomical Mega-Science (as well as the National Key R\&D Program of China with No. 2017YFA0402700). Additional funding support is provided by the Science and Technology Facilities Council of the United Kingdom and participating universities and organizations in the United Kingdom and Canada. Additional funds for the construction of SCUBA-2 were provided by the Canada Foundation for Innovation.

This paper makes use of the following ALMA data: ADS/JAO.ALMA\#2017.1.00629.S. ALMA is a partnership of ESO (representing its member states), NSF (USA) and NINS (Japan), together with NRC (Canada), NSC and ASIAA (Taiwan), and KASI (Republic of Korea), in cooperation with the Republic of Chile. The Joint ALMA Observatory is operated by ESO, AUI/NRAO and NAOJ.

For the majority of the time over which this research was carried out, T.R. was supported by the Science and Technology Facilities Council (STFC) through grant ST/R504725/1. T.R. also thanks the Waterloo Centre for Astrophysics and generous funding to Brian McNamara from the Canadian Space Agency and the National Science and Engineering Research Council of Canada. A.C.E. acknowledges support from STFC grant ST/P00541/1. JWK acknowledges support from the National Research Foundation of Korea (NRF), grant No. NRF-2019R1C1C1002796, funded by the Korean government (MSIT). S.K. acknowledges support from the European Research Council (ERC) under the European Unions Horizon 2020 research and innovation programme under grant agreement No.~771282. A.C. and J.B. acknowledge support from the National Research Foundation of Korea (NRF), grant No. 2018R1D1A1B07048314.

This research made use of \astropy{} \citep{the_astropy_collaboration_astropy_2013,the_astropy_collaboration_astropy_2018}, \matplotlib{} \citep{hunter_matplotlib_2007}, \numpy{} \citep{walt_numpy_2011,harris_array_2020}, \python{} \citep{van_rossum_python_2009}, and \scipy{} \citep{jones_scipy_2011,virtanen_scipy_2020}. We thank their developers for maintaining them and making them freely available.

This publication makes use of data obtained at Mets\"ahovi Radio Observatory, operated by Aalto University in Finland.

\section*{Data Availability}

The data underlying this article will be shared on reasonable request to the corresponding author.

\appendix
\clearpage
\setcounter{page}{1}
\section{Masking of Lightcurves}
\label{sec:MaskingAppendix}
\begin{figure*}
	\includegraphics[width=1\textwidth]{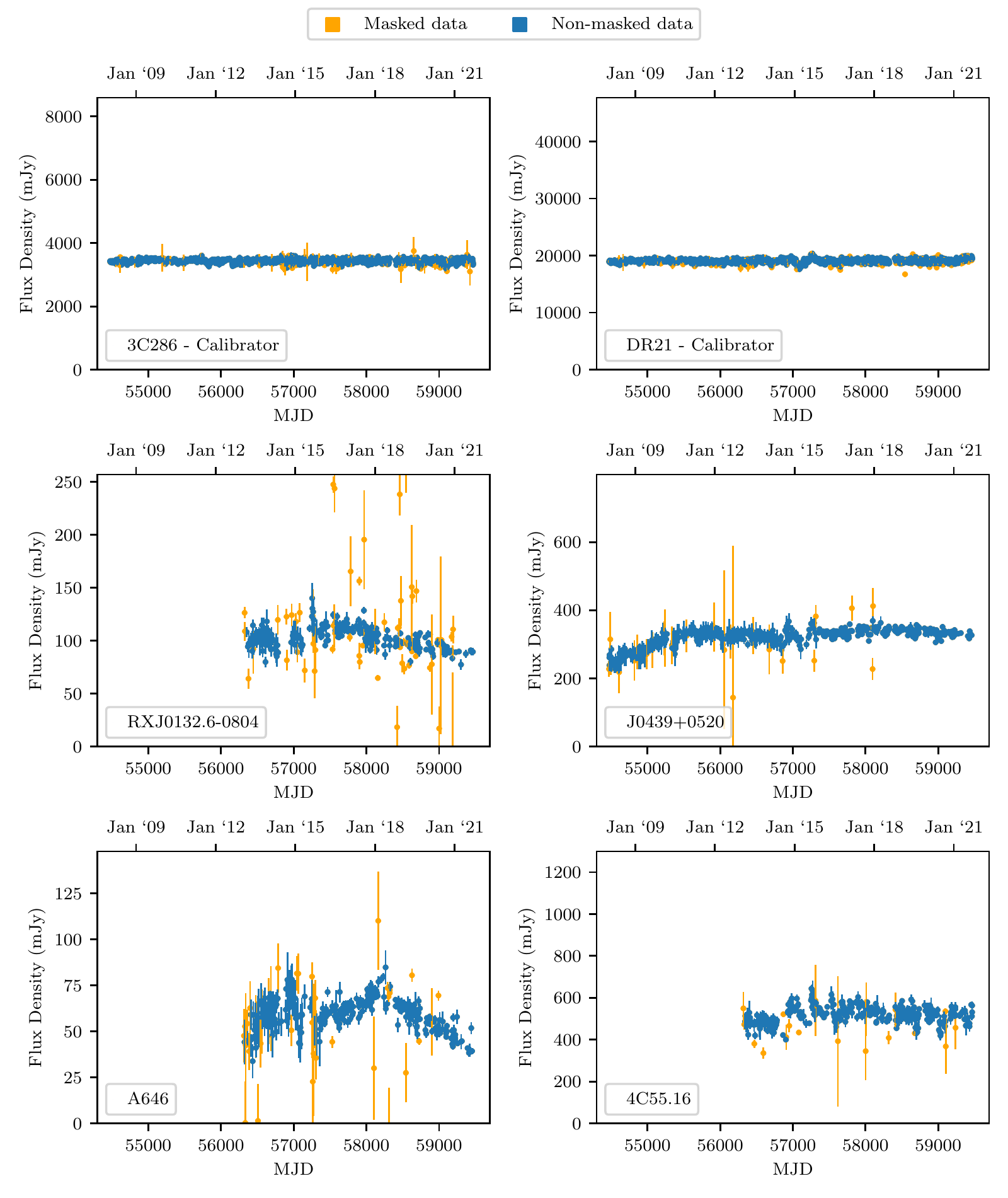}
     \caption{The 15~GHz OVRO spectra shown in Fig \ref{fig:combined_lightcurves_1}, with masked data shown in orange.}
    \label{fig:simulation_lightcurves_1}
\end{figure*}

\begin{figure*}
	\includegraphics[width=1\textwidth]{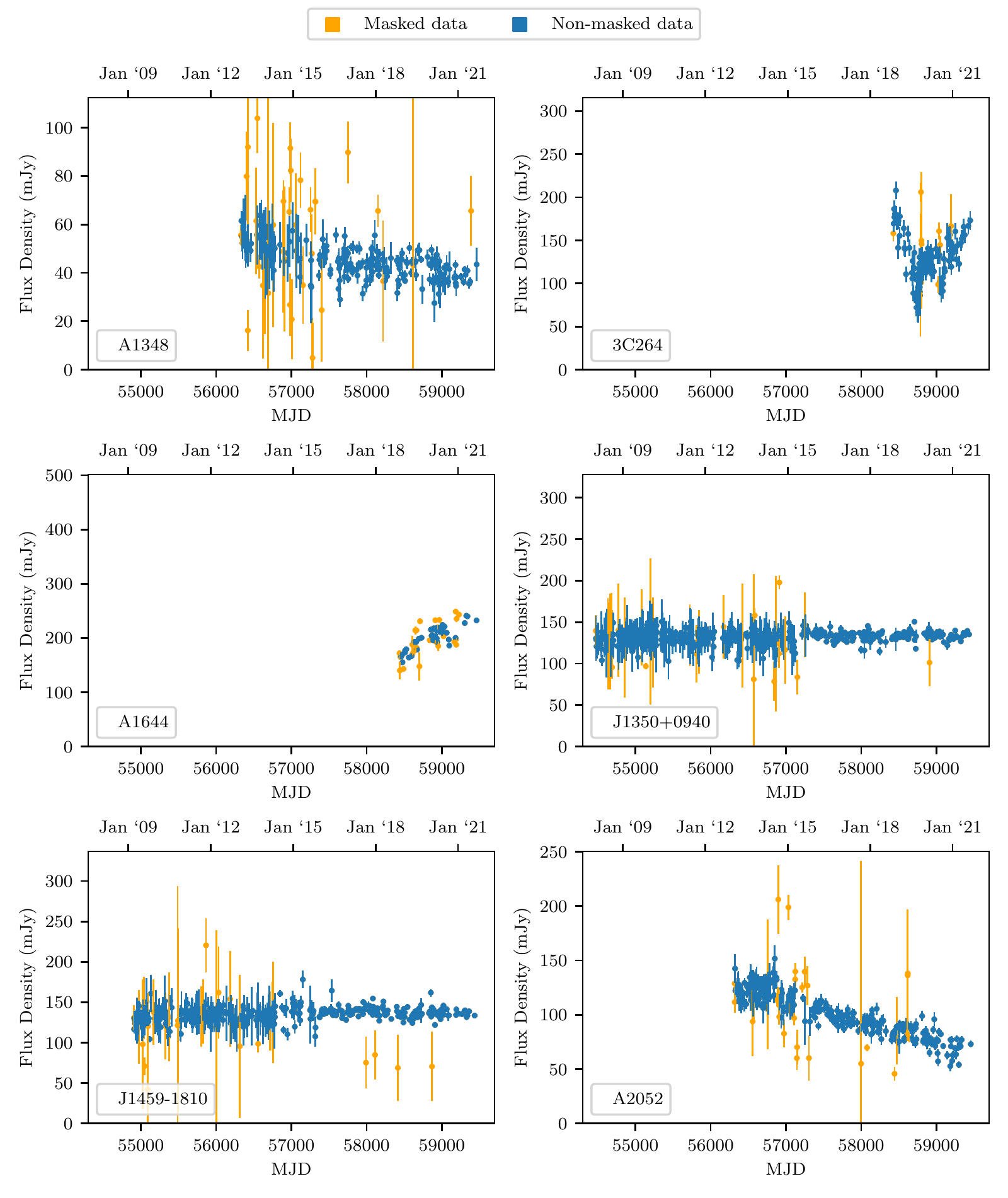}
     \caption{The 15~GHz OVRO spectra shown in Fig \ref{fig:combined_lightcurves_2}, with masked data shown in orange.}
    \label{fig:simulation_lightcurves_2}
\end{figure*}

\begin{figure*}
	\includegraphics[width=1\textwidth]{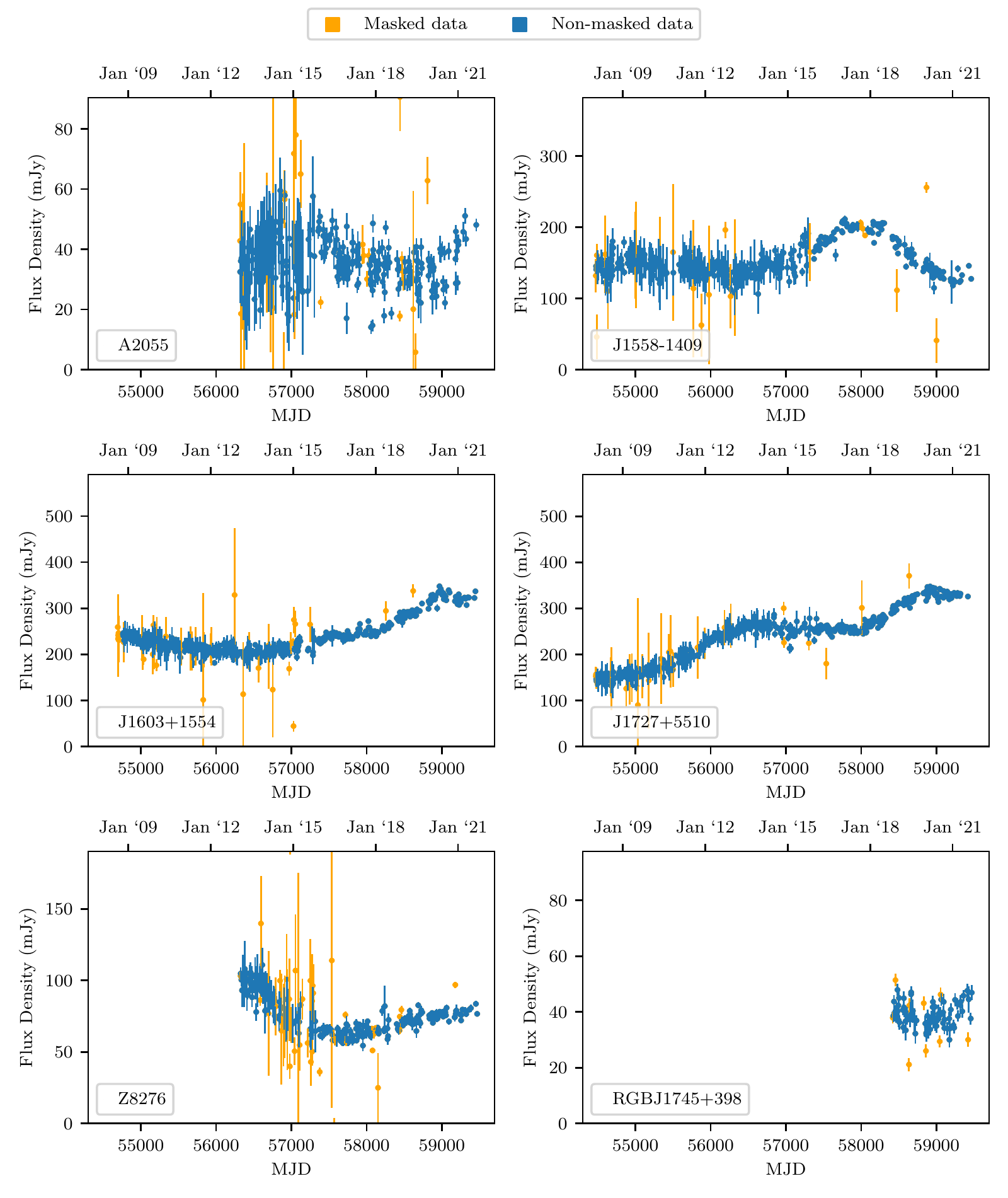}
     \caption{The 15~GHz OVRO spectra shown in Fig \ref{fig:combined_lightcurves_3}, with masked data shown in orange.}
    \label{fig:simulation_lightcurves_3}
\end{figure*}

\begin{figure*}
	\includegraphics[width=1\textwidth]{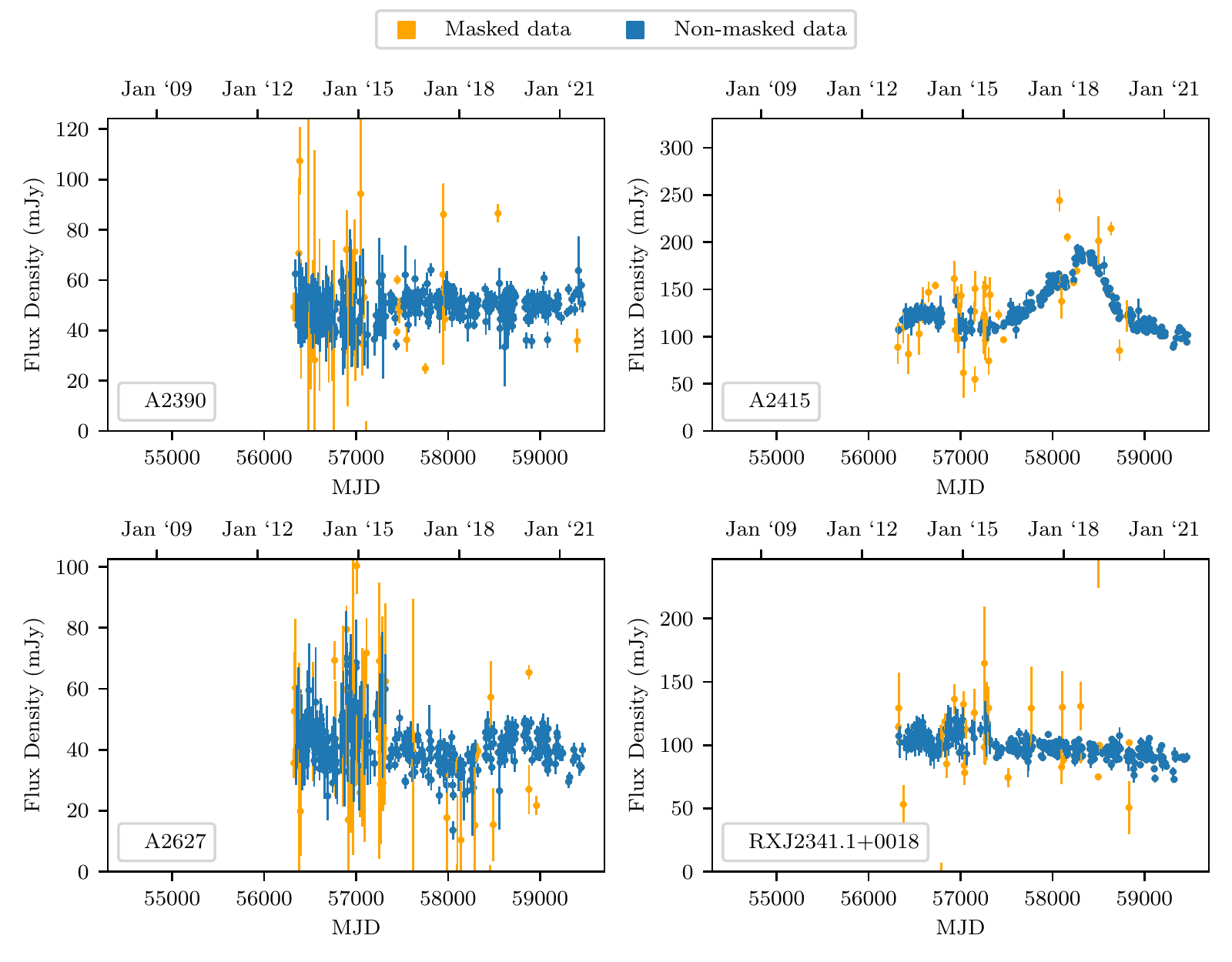}
     \caption{The 15~GHz OVRO spectra shown in Fig \ref{fig:combined_lightcurves_4}, with masked data shown in orange.}
    \label{fig:simulation_lightcurves_4}
\end{figure*}

\section{Mock Lightcurves}
\label{sec:MockSpectraAppendix}
\begin{figure*}
	\includegraphics[width=1\textwidth]{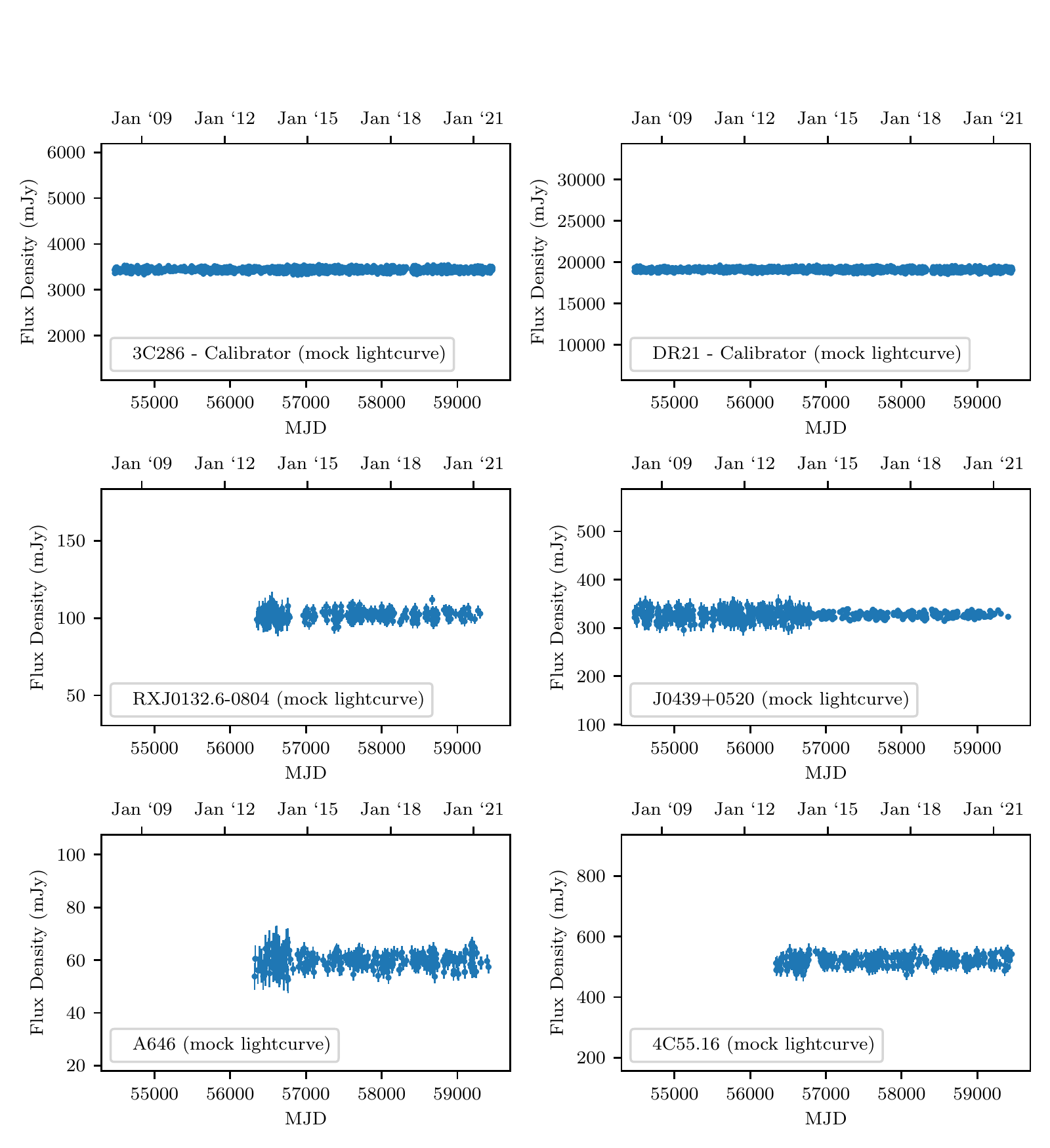}
     \caption{Example mock lightcurves produced along the lines described in \S\ref{sec:mock_spectra}.}
    \label{fig:simulation_lightcurves_1}
\end{figure*}

\begin{figure*}
	\includegraphics[width=1\textwidth]{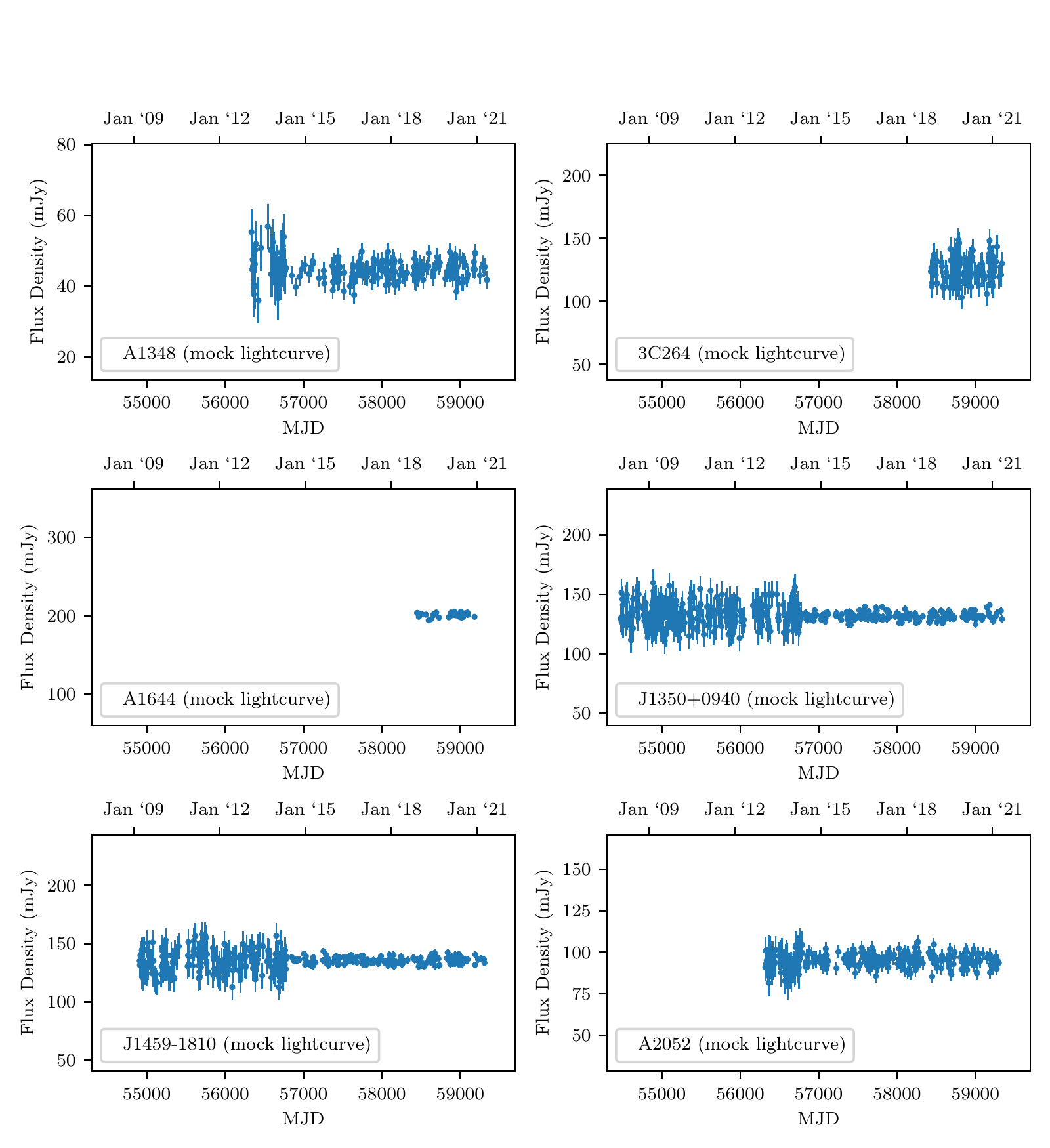}
     \caption{Example mock lightcurves produced along the lines described in \S\ref{sec:mock_spectra}.}
    \label{fig:simulation_lightcurves_2}
\end{figure*}

\begin{figure*}
	\includegraphics[width=1\textwidth]{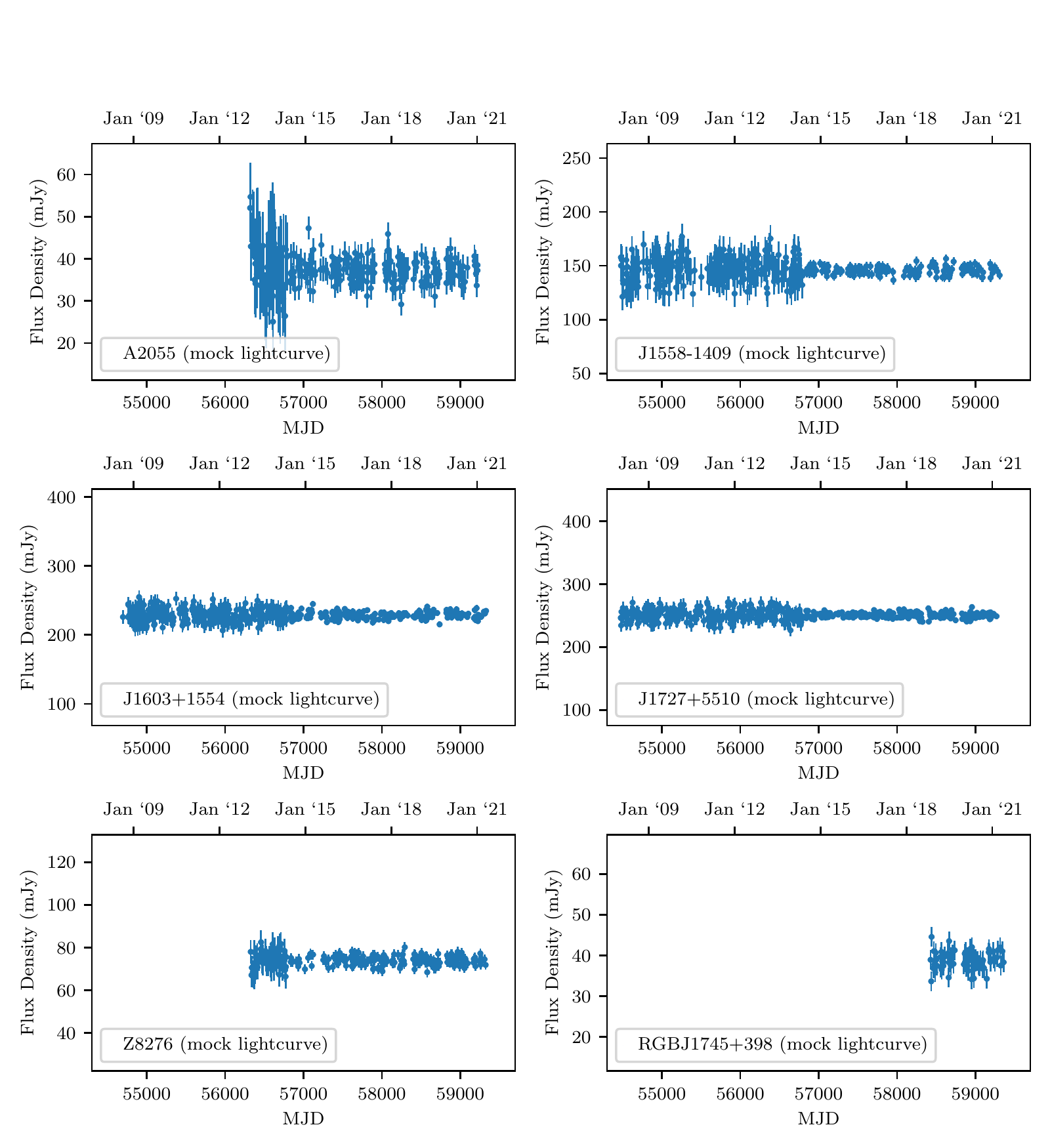}
     \caption{Example mock lightcurves produced along the lines described in \S\ref{sec:mock_spectra}.}
    \label{fig:simulation_lightcurves_3}
\end{figure*}

\begin{figure*}
	\includegraphics[width=1\textwidth]{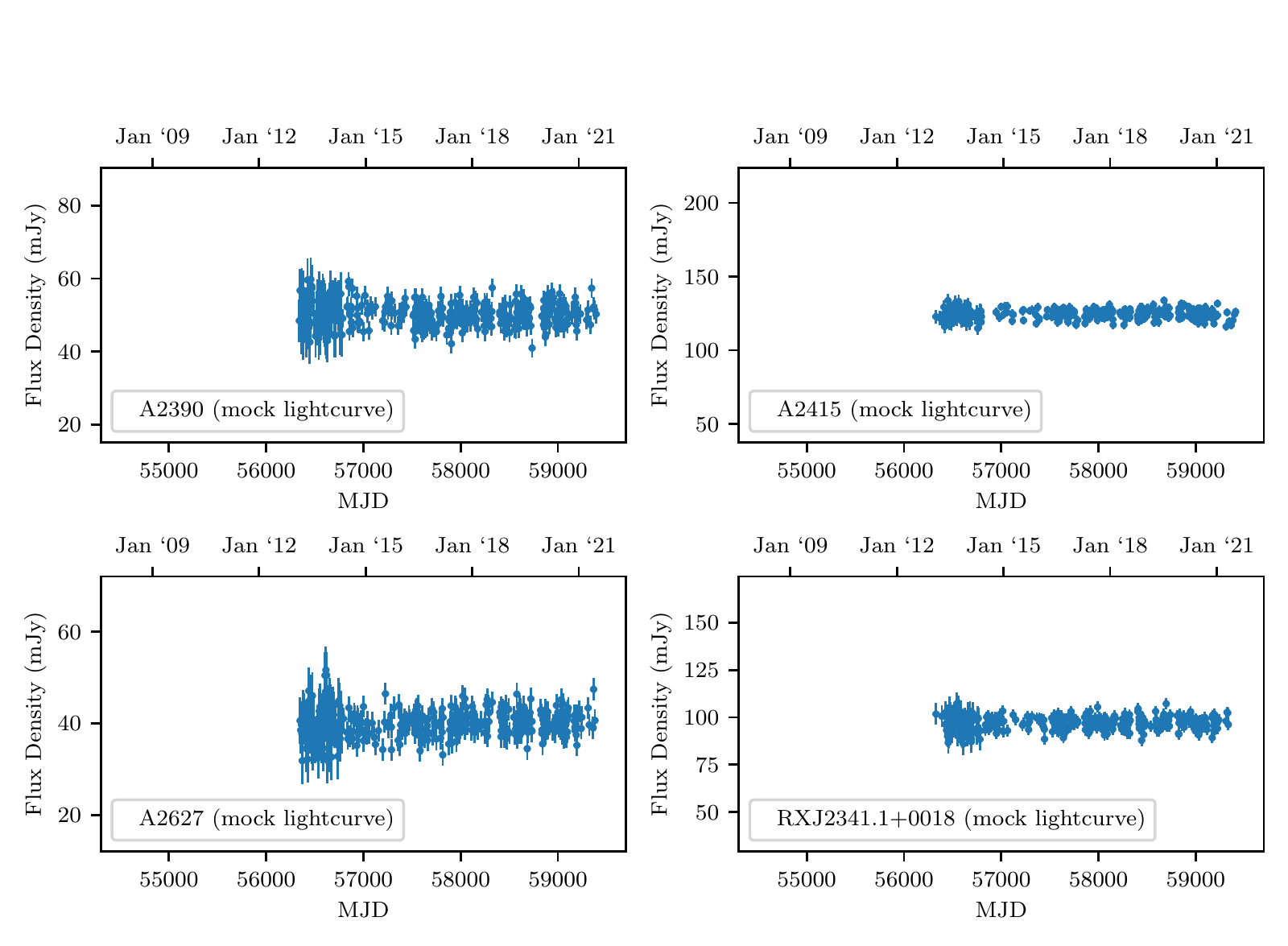}
     \caption{Example mock lightcurves produced along the lines described in \S\ref{sec:mock_spectra}.}
    \label{fig:simulation_lightcurves_4}
\end{figure*}

\section{NIKA2 Data}
\label{sec:NIKA2Appendix}

\begin{table*}
	\caption{150 GHz NIKA2 flux densities, and the spectral indices calculated with the nearest 15 GHz OVRO 40m observations. We also include NIKA2 flux densities of several core-dominated BCGs not included in the OVRO BCG sample.}
	\centering
	\begin{tabular}{lcccccc}
\hline
Source & NIKA2 Obs. & NIKA2 flux density & OVRO Obs. & OVRO flux density & Time Difference & $\alpha$ \\
 & mjd (ISO) & / mJy & mjd (ISO) & / mJy & / days & \\
\hline
J0439+0520  &  58549 (2019-03-07)  &  75 $\pm$ 1  &  58565 (2019-03-22)  &  342 $\pm$ 4  &  16  &  -0.66 $\pm$ 0.0 \\
J0439+0520  &  58786 (2019-10-30)  &  89 $\pm$ 5  &  58811 (2019-11-24)  &  347 $\pm$ 5  &  25  &  -0.59 $\pm$ 0.02 \\
J0439+0520  &  58791 (2019-11-04)  &  49 $\pm$ 1  &  58811 (2019-11-24)  &  347 $\pm$ 5  &  20  &  -0.85 $\pm$ 0.01 \\
4C55.16  &  58550 (2019-03-08)  &  86 $\pm$ 1  &  58546 (2019-03-04)  &  523 $\pm$ 21  &  4  &  -0.79 $\pm$ -0.01 \\
4C55.16  &  58828 (2019-12-11)  &  74 $\pm$ 1  &  58827 (2019-12-10)  &  562 $\pm$ 13  &  1  &  -0.87 $\pm$ -0.01 \\
A1348  &  58922 (2020-03-14)  &  8 $\pm$ 1  &  58926 (2020-03-18)  &  38 $\pm$ 2  &  4  &  -0.7 $\pm$ 0.02 \\
3C264  &  58550 (2019-03-08)  &  125 $\pm$ 1  &  58567 (2019-03-25)  &  166 $\pm$ 10  &  17  &  -0.13 $\pm$ -0.02 \\
3C264  &  58828 (2019-12-11)  &  111 $\pm$ 1  &  58825 (2019-12-08)  &  117 $\pm$ 9  &  3  &  -0.05 $\pm$ -0.02 \\
A1644  &  58553 (2019-03-11)  &  59 $\pm$ 1  &  58561 (2019-03-19)  &  164 $\pm$ 3  &  8  &  -0.46 $\pm$ 0.0 \\
A1644  &  58897 (2020-02-18)  &  77 $\pm$ 1  &  58893 (2020-02-13)  &  218 $\pm$ 3  &  4  &  -0.43 $\pm$ 0.0 \\
J1350+0940  &  58565 (2019-03-23)  &  18 $\pm$ 1  &  58567 (2019-03-25)  &  135 $\pm$ 3  &  2  &  -0.87 $\pm$ 0.01 \\
J1350+0940  &  58897 (2020-02-18)  &  18 $\pm$ 1  &  58895 (2020-02-16)  &  136 $\pm$ 2  &  2  &  -0.87 $\pm$ 0.02 \\
J1459-1810  &  58567 (2019-03-25)  &  21 $\pm$ 1  &  58572 (2019-03-30)  &  128 $\pm$ 2  &  5  &  -0.81 $\pm$ 0.01 \\
A2052  &  58550 (2019-03-08)  &  44 $\pm$ 1  &  58557 (2019-03-15)  &  88 $\pm$ 4  &  7  &  -0.3 $\pm$ -0.01 \\
A2052  &  58829 (2019-12-12)  &  37 $\pm$ 1  &  58830 (2019-12-12)  &  75 $\pm$ 3  &  1  &  -0.31 $\pm$ -0.01 \\
A2055  &  58565 (2019-03-23)  &  30 $\pm$ 1  &  58566 (2019-03-24)  &  30 $\pm$ 2  &  1  &  -0.06 $\pm$ -0.01 \\
A2055  &  58831 (2019-12-14)  &  24 $\pm$ 1  &  58830 (2019-12-12)  &  37 $\pm$ 2  &  1  &  -0.18 $\pm$ -0.01 \\
J1558-1409  &  58553 (2019-03-11)  &  47 $\pm$ 1  &  58577 (2019-04-04)  &  172 $\pm$ 13  &  24  &  -0.57 $\pm$ -0.01 \\
J1558-1409  &  58897 (2020-02-18)  &  32 $\pm$ 2  &  58894 (2020-02-14)  &  142 $\pm$ 3  &  3  &  -0.65 $\pm$ 0.01 \\
J1603+1554  &  58553 (2019-03-11)  &  84 $\pm$ 1  &  58557 (2019-03-15)  &  290 $\pm$ 4  &  4  &  -0.54 $\pm$ -0.0 \\
J1603+1554  &  58829 (2019-12-12)  &  79 $\pm$ 1  &  58825 (2019-12-07)  &  315 $\pm$ 6  &  4  &  -0.59 $\pm$ -0.0 \\
J1727+5510  &  58550 (2019-03-08)  &  89 $\pm$ 1  &  58555 (2019-03-12)  &  309 $\pm$ 5  &  5  &  -0.54 $\pm$ -0.0 \\
J1727+5510  &  58790 (2019-11-03)  &  64 $\pm$ 2  &  58829 (2019-12-11)  &  333 $\pm$ 5  &  39  &  -0.71 $\pm$ 0.01 \\
Z8276  &  58554 (2019-03-12)  &  36 $\pm$ 1  &  58557 (2019-03-15)  &  73 $\pm$ 2  &  3  &  -0.3 $\pm$ -0.0 \\
Z8276  &  58831 (2019-12-14)  &  30 $\pm$ 1  &  58832 (2019-12-14)  &  76 $\pm$ 2  &  1  &  -0.41 $\pm$ 0.0 \\
RGBJ1745+398  &  58554 (2019-03-12)  &  30 $\pm$ 1  &  58555 (2019-03-12)  &  42 $\pm$ 2  &  1  &  -0.15 $\pm$ -0.01 \\
RGBJ1745+398  &  58790 (2019-11-03)  &  23 $\pm$ 1  &  58735 (2019-09-09)  &  38 $\pm$ 2  &  55  &  -0.21 $\pm$ -0.01 \\
A2627  &  58922 (2020-03-14)  &  25 $\pm$ 1  &  58913 (2020-03-04)  &  37 $\pm$ 2  &  9  &  -0.23 $\pm$ -0.01 \\
\hline
NGC5044 & 58553 (2019-03-11)& 37.4 $\pm$ 0.9 & & & & \\
NGC5044 & 58897 (2020-02-18)& 35.4 $\pm$ 1.3 & & & & \\
A496 & 58550 (2019-03-08)& 21.0 $\pm$ 1.0 & & & & \\
A496 & 58831 (2019-12-14)& 12.4 $\pm$ 0.9 & & & & \\
A1795 & 58829 (2019-12-12)& 4.7 $\pm$ 0.6 & & & & \\
A3581 & 58553 (2019-03-11)& 99.9 $\pm$ 1.7 & & & & \\
E1821+64 & 58790 (2019-11-03)& 9.1 $\pm$ 1.0 & & & & \\
NGC6338 & 58565 (2019-03-23)& 4.2 $\pm$ 0.7 & & & & \\
NGC6338 & 58790 (2019-11-03)& 4.9 $\pm$ 0.8 & & & & \\
R1504-02 & 58565 (2019-03-23)& 5.7 $\pm$ 0.7 & & & & \\
R1504-02 & 58828 (2019-12-11)& 16.8 $\pm$ 2.0 & & & & \\
R1504-02 & 58831 (2019-12-14)& 4.0 $\pm$ 0.6 & & & & \\
R1832+68 & 58790 (2019-11-03)& 6.1 $\pm$ 0.9 & & & & \\
Z8193 & 58554 (2019-03-12)& 28.5 $\pm$ 0.6 & & & & \\
Z8193 & 58790 (2019-11-03)& 17.0 $\pm$ 1.8 & & & & \\
Z8193 & 58831 (2019-12-14)& 21.1 $\pm$ 0.8 & & & & \\
\hline
	\end{tabular}
	\label{tab:NIKA2_obs1}
\end{table*}

\section{KVN Data}
\label{sec:KVNAppendix}

\begin{table*}
	\caption{Contemporaneous 22 and 43 GHz KVN flux densities and their spectral indices. Continued on the next page.}
	\centering
	\begin{tabular}{lcccc}
\hline
Source & mjd (ISO) & 22 GHz flux density / mJy & 43 GHz flux density / mJy & $\alpha$ \\
\hline
RXJ0132.6-0804 & 58534 (2019-02-20) & 74.2 $\pm$ 7.9 & 52.4 $\pm$ 4.0 & -0.5 $\pm$ 0.2 \\
RXJ0132.6-0804 & 58584 (2019-04-11) & 75.0 $\pm$ 6.4 & 51.2 $\pm$ 3.6 & -0.6 $\pm$ 0.1 \\
RXJ0132.6-0804 & 58827 (2019-12-10) & 77.7 $\pm$ 5.9 & 52.0 $\pm$ 4.4 & -0.6 $\pm$ 0.2 \\
RXJ0132.6-0804 & 58975 (2020-05-06) & 69.3 $\pm$ 7.0 & 43.6 $\pm$ 3.2 & -0.7 $\pm$ 0.2 \\
RXJ0132.6-0804 & 59006 (2020-06-06) & 72.6 $\pm$ 6.3 & 46.2 $\pm$ 3.9 & -0.7 $\pm$ 0.2 \\
RXJ0132.6-0804 & 59182 (2020-11-29) & 79.4 $\pm$ 9.2 & 49.1 $\pm$ 9.3 & -0.7 $\pm$ 0.3 \\
J0439+0520 & 58421 (2018-10-30) & 286.0 $\pm$ 16.1 & 139.0 $\pm$ 11.3 & -1.1 $\pm$ 0.2 \\
J0439+0520 & 58450 (2018-11-28) & 277.0 $\pm$ 17.1 & 158.0 $\pm$ 9.6 & -0.8 $\pm$ 0.2 \\
J0439+0520 & 58534 (2019-02-20) & 271.0 $\pm$ 18.0 & 146.0 $\pm$ 17.5 & -0.9 $\pm$ 0.2 \\
J0439+0520 & 58584 (2019-04-11) & 261.0 $\pm$ 16.0 & 121.0 $\pm$ 8.5 & -1.1 $\pm$ 0.2 \\
J0439+0520 & 58827 (2019-12-10) & 267.0 $\pm$ 15.7 & 138.0 $\pm$ 11.2 & -1.0 $\pm$ 0.2 \\
J0439+0520 & 58975 (2020-05-06) & 247.0 $\pm$ 15.4 & 121.0 $\pm$ 7.6 & -1.1 $\pm$ 0.2 \\
J0439+0520 & 59006 (2020-06-06) & 247.0 $\pm$ 14.7 & 120.0 $\pm$ 8.2 & -1.1 $\pm$ 0.2 \\
J0439+0520 & 59182 (2020-11-29) & 249.0 $\pm$ 17.5 & 159.0 $\pm$ 9.4 & -0.7 $\pm$ 0.2 \\
A646 & 58421 (2018-10-30) & 54.0 $\pm$ 3.6 & 36.1 $\pm$ 3.3 & -0.6 $\pm$ 0.2 \\
A646 & 58450 (2018-11-28) & 52.7 $\pm$ 3.4 & 37.3 $\pm$ 3.0 & -0.5 $\pm$ 0.1 \\
A646 & 58512 (2019-01-29) & 52.2 $\pm$ 3.9 & 33.8 $\pm$ 3.0 & -0.6 $\pm$ 0.2 \\
A646 & 58847 (2019-12-30) & 43.3 $\pm$ 2.7 & -- & -- \\
A646 & 58974 (2020-05-05) & 49.9 $\pm$ 4.4 & 35.3 $\pm$ 3.1 & -0.5 $\pm$ 0.2 \\
A646 & 59182 (2020-11-29) & 36.0 $\pm$ 2.4 & -- & -- \\
4C55.16 & 58421 (2018-10-30) & 197.0 $\pm$ 22.6 & 47.0 $\pm$ 5.2 & -2.1 $\pm$ 0.5 \\
4C55.16 & 58450 (2018-11-28) & 225.0 $\pm$ 26.4 & 35.6 $\pm$ 4.3 & -2.8 $\pm$ 0.6 \\
4C55.16 & 58512 (2019-01-29) & 203.0 $\pm$ 23.8 & 35.1 $\pm$ 4.3 & -2.6 $\pm$ 0.6 \\
4C55.16 & 58573 (2019-03-31) & 220.0 $\pm$ 27.1 & 41.9 $\pm$ 5.8 & -2.5 $\pm$ 0.5 \\
4C55.16 & 58847 (2019-12-30) & 190.0 $\pm$ 23.6 & 39.1 $\pm$ 4.3 & -2.4 $\pm$ 0.5 \\
4C55.16 & 58974 (2020-05-05) & 211.0 $\pm$ 24.5 & 50.9 $\pm$ 5.1 & -2.1 $\pm$ 0.5 \\
4C55.16 & 59182 (2020-11-29) & 208.0 $\pm$ 23.6 & 47.8 $\pm$ 5.7 & -2.2 $\pm$ 0.5 \\
A1348 & 58421 (2018-10-30) & 32.3 $\pm$ 2.7 & -- & -- \\
A1348 & 58450 (2018-11-28) & 24.7 $\pm$ 2.4 & -- & -- \\
A1348 & 58512 (2019-01-29) & 29.0 $\pm$ 2.5 & -- & -- \\
3C264 & 58512 (2019-01-29) & 136.0 $\pm$ 8.4 & 103.0 $\pm$ 7.2 & -0.4 $\pm$ 0.1 \\
3C264 & 58573 (2019-03-31) & 110.0 $\pm$ 7.4 & 77.5 $\pm$ 5.6 & -0.5 $\pm$ 0.1 \\
3C264 & 58847 (2019-12-30) & 87.3 $\pm$ 6.2 & 62.0 $\pm$ 4.8 & -0.5 $\pm$ 0.1 \\
3C264 & 58852 (2020-01-04) & 90.4 $\pm$ 6.7 & 69.5 $\pm$ 4.6 & -0.4 $\pm$ 0.1 \\
3C264 & 58974 (2020-05-05) & 107.0 $\pm$ 7.8 & 75.4 $\pm$ 5.4 & -0.5 $\pm$ 0.1 \\
3C264 & 58988 (2020-05-19) & 102.0 $\pm$ 7.2 & 79.3 $\pm$ 5.7 & -0.4 $\pm$ 0.1 \\
3C264 & 59182 (2020-11-29) & 106.0 $\pm$ 7.0 & 79.9 $\pm$ 4.9 & -0.4 $\pm$ 0.1 \\
A1644 & 58450 (2018-11-28) & 140.0 $\pm$ 10.0 & 71.3 $\pm$ 6.2 & -1.0 $\pm$ 0.2 \\
A1644 & 58512 (2019-01-29) & 158.0 $\pm$ 11.1 & 75.2 $\pm$ 7.1 & -1.1 $\pm$ 0.3 \\
A1644 & 58573 (2019-03-31) & 161.0 $\pm$ 10.6 & 81.7 $\pm$ 6.8 & -1.0 $\pm$ 0.2 \\
A1644 & 58847 (2019-12-30) & 177.0 $\pm$ 10.4 & 92.5 $\pm$ 7.7 & -1.0 $\pm$ 0.2 \\
A1644 & 58852 (2020-01-04) & 181.0 $\pm$ 11.5 & 103.0 $\pm$ 8.2 & -0.8 $\pm$ 0.2 \\
A1644 & 58974 (2020-05-05) & 195.0 $\pm$ 12.4 & 100.0 $\pm$ 8.3 & -1.0 $\pm$ 0.2 \\
A1644 & 58988 (2020-05-19) & 191.0 $\pm$ 10.9 & 99.4 $\pm$ 8.1 & -1.0 $\pm$ 0.2 \\
A1644 & 59182 (2020-11-29) & 169.0 $\pm$ 10.7 & 86.0 $\pm$ 7.5 & -1.0 $\pm$ 0.2 \\
J1350+0940 & 58421 (2018-10-30) & 88.9 $\pm$ 5.6 & 53.1 $\pm$ 5.0 & -0.8 $\pm$ 0.2 \\
J1350+0940 & 58450 (2018-11-28) & 86.9 $\pm$ 5.3 & 42.3 $\pm$ 3.8 & -1.1 $\pm$ 0.2 \\
J1350+0940 & 58512 (2019-01-29) & 87.1 $\pm$ 5.1 & 41.5 $\pm$ 2.9 & -1.1 $\pm$ 0.2 \\
J1350+0940 & 58843 (2019-12-26) & 86.5 $\pm$ 5.3 & 40.5 $\pm$ 2.9 & -1.1 $\pm$ 0.2 \\
J1350+0940 & 58852 (2020-01-04) & 85.4 $\pm$ 5.2 & 40.3 $\pm$ 2.7 & -1.1 $\pm$ 0.2 \\
J1350+0940 & 58988 (2020-05-19) & 89.9 $\pm$ 5.5 & 43.4 $\pm$ 3.6 & -1.1 $\pm$ 0.2 \\
J1459-1810 & 58421 (2018-10-30) & 107.0 $\pm$ 6.1 & 62.4 $\pm$ 7.9 & -0.8 $\pm$ 0.2 \\
J1459-1810 & 58514 (2019-01-31) & 100.0 $\pm$ 6.3 & 52.5 $\pm$ 3.8 & -1.0 $\pm$ 0.2 \\
J1459-1810 & 58843 (2019-12-26) & 98.4 $\pm$ 6.1 & 55.5 $\pm$ 3.3 & -0.9 $\pm$ 0.2 \\
J1459-1810 & 58969 (2020-04-30) & 99.9 $\pm$ 6.2 & 56.7 $\pm$ 4.1 & -0.8 $\pm$ 0.2 \\
J1459-1810 & 59182 (2020-11-29) & 96.5 $\pm$ 6.7 & 53.7 $\pm$ 3.8 & -0.9 $\pm$ 0.2 \\
A2052 & 58421 (2018-10-30) & 65.3 $\pm$ 8.0 & 43.1 $\pm$ 6.9 & -0.6 $\pm$ 0.2 \\
A2052 & 58514 (2019-01-31) & 67.8 $\pm$ 8.5 & 42.7 $\pm$ 4.5 & -0.7 $\pm$ 0.2 \\
A2052 & 58573 (2019-03-31) & 71.4 $\pm$ 11.1 & 49.1 $\pm$ 4.1 & -0.6 $\pm$ 0.2 \\
A2052 & 58843 (2019-12-26) & 64.8 $\pm$ 6.4 & 44.8 $\pm$ 3.2 & -0.6 $\pm$ 0.2 \\
A2052 & 58969 (2020-04-30) & 57.5 $\pm$ 5.6 & -- & -- \\
A2052 & 59182 (2020-11-29) & 54.6 $\pm$ 4.6 & 42.3 $\pm$ 4.1 & -0.4 $\pm$ 0.1 \\
\hline
	\end{tabular}
	\label{tab:KVN_obs1}
\end{table*}

\begin{table*}
	\centering
	\begin{tabular}{lcccc}
\hline
Source & mjd (ISO) & 22 GHz flux density / mJy & 43 GHz flux density / mJy & $\alpha$ \\
\hline
A2055 & 58421 (2018-10-30) & 31.2 $\pm$ 2.7 & -- & -- \\
A2055 & 58514 (2019-01-31) & 31.9 $\pm$ 2.5 & 28.5 $\pm$ 2.2 & -0.2 $\pm$ 0.1 \\
A2055 & 58843 (2019-12-26) & 31.3 $\pm$ 2.2 & 29.6 $\pm$ 2.6 & -0.1 $\pm$ 0.1 \\
J1558-1409 & 58421 (2018-10-30) & 139.0 $\pm$ 14.4 & 56.6 $\pm$ 9.2 & -1.3 $\pm$ 0.3 \\
J1558-1409 & 58514 (2019-01-31) & 122.0 $\pm$ 16.1 & 63.7 $\pm$ 10.9 & -1.0 $\pm$ 0.3 \\
J1558-1409 & 58573 (2019-03-31) & 117.0 $\pm$ 8.8 & 75.2 $\pm$ 5.6 & -0.7 $\pm$ 0.2 \\
J1558-1409 & 58843 (2019-12-26) & 92.4 $\pm$ 7.0 & 48.9 $\pm$ 3.5 & -0.9 $\pm$ 0.2 \\
J1558-1409 & 58852 (2020-01-04) & 97.5 $\pm$ 8.4 & 48.3 $\pm$ 4.8 & -1.0 $\pm$ 0.2 \\
J1558-1409 & 58969 (2020-04-30) & 93.2 $\pm$ 8.2 & -- & -- \\
J1558-1409 & 58988 (2020-05-19) & 91.1 $\pm$ 6.9 & 49.6 $\pm$ 3.9 & -0.9 $\pm$ 0.2 \\
J1558-1409 & 59182 (2020-11-29) & 83.5 $\pm$ 7.5 & -- & -- \\
J1603+1554 & 58421 (2018-10-30) & 262.0 $\pm$ 15.1 & 155.0 $\pm$ 13.0 & -0.8 $\pm$ 0.2 \\
J1603+1554 & 58514 (2019-01-31) & 264.0 $\pm$ 16.2 & 172.0 $\pm$ 10.2 & -0.6 $\pm$ 0.1 \\
J1603+1554 & 58573 (2019-03-31) & 270.0 $\pm$ 15.5 & 173.0 $\pm$ 9.5 & -0.7 $\pm$ 0.2 \\
J1603+1554 & 58843 (2019-12-26) & 291.0 $\pm$ 16.8 & 182.0 $\pm$ 10.6 & -0.7 $\pm$ 0.2 \\
J1603+1554 & 58852 (2020-01-04) & 289.0 $\pm$ 17.5 & 175.0 $\pm$ 10.6 & -0.7 $\pm$ 0.2 \\
J1603+1554 & 58969 (2020-04-30) & 308.0 $\pm$ 20.2 & 183.0 $\pm$ 11.4 & -0.8 $\pm$ 0.2 \\
J1603+1554 & 58988 (2020-05-19) & 320.0 $\pm$ 18.3 & 207.0 $\pm$ 11.8 & -0.6 $\pm$ 0.2 \\
J1603+1554 & 59182 (2020-11-29) & 269.0 $\pm$ 16.3 & 166.0 $\pm$ 9.5 & -0.7 $\pm$ 0.2 \\
J1727+5510 & 58421 (2018-10-30) & 271.0 $\pm$ 17.0 & 156.0 $\pm$ 10.6 & -0.8 $\pm$ 0.2 \\
J1727+5510 & 58514 (2019-01-31) & 282.0 $\pm$ 16.9 & 176.0 $\pm$ 10.3 & -0.7 $\pm$ 0.2 \\
J1727+5510 & 58573 (2019-03-31) & 289.0 $\pm$ 16.5 & 187.0 $\pm$ 10.8 & -0.6 $\pm$ 0.2 \\
J1727+5510 & 58852 (2020-01-04) & 295.0 $\pm$ 17.4 & 178.0 $\pm$ 11.0 & -0.8 $\pm$ 0.2 \\
J1727+5510 & 58969 (2020-04-30) & 307.0 $\pm$ 19.0 & 194.0 $\pm$ 12.6 & -0.7 $\pm$ 0.2 \\
J1727+5510 & 58988 (2020-05-19) & 318.0 $\pm$ 19.1 & 211.0 $\pm$ 12.0 & -0.6 $\pm$ 0.1 \\
J1727+5510 & 59182 (2020-11-29) & 274.0 $\pm$ 15.6 & 169.0 $\pm$ 9.6 & -0.7 $\pm$ 0.2 \\
Z8276 & 58421 (2018-10-30) & 58.4 $\pm$ 3.5 & 37.3 $\pm$ 3.8 & -0.7 $\pm$ 0.2 \\
Z8276 & 58514 (2019-01-31) & 59.0 $\pm$ 4.4 & 37.9 $\pm$ 2.9 & -0.7 $\pm$ 0.2 \\
Z8276 & 58573 (2019-03-31) & 59.7 $\pm$ 3.8 & 43.6 $\pm$ 3.7 & -0.5 $\pm$ 0.1 \\
Z8276 & 58852 (2020-01-04) & 58.3 $\pm$ 4.0 & -- & -- \\
Z8276 & 58969 (2020-04-30) & 58.1 $\pm$ 3.8 & -- & -- \\
Z8276 & 58975 (2020-05-06) & 63.4 $\pm$ 4.8 & 44.9 $\pm$ 4.4 & -0.5 $\pm$ 0.2 \\
Z8276 & 59006 (2020-06-06) & 64.1 $\pm$ 5.7 & 49.8 $\pm$ 7.0 & -0.4 $\pm$ 0.2 \\
Z8276 & 59182 (2020-11-29) & 61.1 $\pm$ 3.8 & -- & -- \\
RGBJ1745+398 & 58573 (2019-03-31) & 35.3 $\pm$ 2.7 & 29.7 $\pm$ 2.7 & -0.3 $\pm$ 0.1 \\
RGBJ1745+398 & 58852 (2020-01-04) & 32.0 $\pm$ 2.6 & -- & -- \\
RGBJ1745+398 & 58969 (2020-04-30) & 38.3 $\pm$ 2.8 & -- & -- \\
A2390 & 58421 (2018-10-30) & 45.4 $\pm$ 3.1 & 35.2 $\pm$ 3.5 & -0.4 $\pm$ 0.1 \\
A2390 & 58534 (2019-02-20) & 42.4 $\pm$ 3.0 & -- & -- \\
A2390 & 58827 (2019-12-10) & 45.0 $\pm$ 3.2 & -- & -- \\
A2390 & 58975 (2020-05-06) & 40.7 $\pm$ 3.4 & -- & -- \\
A2390 & 59006 (2020-06-06) & 46.9 $\pm$ 3.5 & -- & -- \\
A2415 & 58421 (2018-10-30) & 161.0 $\pm$ 9.6 & 103.0 $\pm$ 6.6 & -0.7 $\pm$ 0.2 \\
A2415 & 58534 (2019-02-20) & 134.0 $\pm$ 9.3 & 81.4 $\pm$ 5.6 & -0.7 $\pm$ 0.2 \\
A2415 & 58584 (2019-04-11) & 124.0 $\pm$ 7.3 & 70.3 $\pm$ 4.4 & -0.8 $\pm$ 0.2 \\
A2415 & 58827 (2019-12-10) & 92.5 $\pm$ 5.7 & 52.1 $\pm$ 4.0 & -0.9 $\pm$ 0.2 \\
A2415 & 58975 (2020-05-06) & 86.9 $\pm$ 5.4 & 47.4 $\pm$ 4.0 & -0.9 $\pm$ 0.2 \\
A2415 & 59006 (2020-06-06) & 89.7 $\pm$ 5.6 & 60.4 $\pm$ 5.6 & -0.6 $\pm$ 0.2 \\
A2415 & 59182 (2020-11-29) & 76.0 $\pm$ 4.8 & 42.7 $\pm$ 3.1 & -0.9 $\pm$ 0.2 \\
A2627 & 58534 (2019-02-20) & 35.8 $\pm$ 2.9 & -- & -- \\
A2627 & 58827 (2019-12-10) & 39.9 $\pm$ 3.6 & 34.4 $\pm$ 3.0 & -0.2 $\pm$ 0.1 \\
A2627 & 58975 (2020-05-06) & 34.9 $\pm$ 2.8 & -- & -- \\
A2627 & 59006 (2020-06-06) & 46.2 $\pm$ 3.5 & -- & -- \\
RXJ2341.1+0018 & 58421 (2018-10-30) & 75.4 $\pm$ 7.5 & 41.8 $\pm$ 5.7 & -0.9 $\pm$ 0.2 \\
RXJ2341.1+0018 & 58534 (2019-02-20) & 70.4 $\pm$ 10.0 & -- & -- \\
RXJ2341.1+0018 & 58584 (2019-04-11) & 60.8 $\pm$ 7.7 & 36.9 $\pm$ 3.9 & -0.7 $\pm$ 0.2 \\
RXJ2341.1+0018 & 58827 (2019-12-10) & 61.7 $\pm$ 6.5 & -- & -- \\
RXJ2341.1+0018 & 58975 (2020-05-06) & 67.1 $\pm$ 5.1 & 36.8 $\pm$ 4.5 & -0.9 $\pm$ 0.2 \\
RXJ2341.1+0018 & 59006 (2020-06-06) & 69.9 $\pm$ 5.9 & -- & -- \\
\hline
\hline
	\end{tabular}
	\label{tab:KVN_obs2}
\end{table*}

\section{SCUBA2 Data}
\label{sec:SCUBA2Appendix}
\begin{table*}
	\caption{353 GHz SCUBA2 flux densities, and the spectral indices calculated with the nearest 15 GHz OVRO 40m observations. Continued on the next page.}
	\centering
	\begin{tabular}{lcccccc}
\hline
Source & SCUBA2 Obs. & SCUBA2 flux density & OVRO Obs.& OVRO flux density & Time Difference & $\alpha$ \\
 & mjd (ISO) & / mJy & mjd (ISO) & / mJy & / days & \\
\hline
RXJ0132.6-0804  &  56223 (2012-10-23)  &  14 $\pm$ 9  &  56353 (2013-03-01)  &  107 $\pm$ 5  &  130  &  -0.6 $\pm$ 0.15 \\
RXJ0132.6-0804  &  56476 (2013-07-03)  &  21 $\pm$ 8  &  56473 (2013-06-29)  &  103 $\pm$ 6  &  3  &  -0.5 $\pm$ 0.08 \\
RXJ0132.6-0804  &  58628 (2019-05-25)  &  22 $\pm$ 3  &  58618 (2019-05-14)  &  96 $\pm$ 3  &  10  &  -0.46 $\pm$ 0.03 \\
RXJ0132.6-0804  &  58751 (2019-09-25)  &  20 $\pm$ 3  &  58739 (2019-09-13)  &  106 $\pm$ 3  &  12  &  -0.49 $\pm$ 0.03 \\
RXJ0132.6-0804  &  59021 (2020-06-21)  &  30 $\pm$ 2  &  59042 (2020-07-11)  &  97 $\pm$ 4  &  21  &  -0.35 $\pm$ 0.01 \\
RXJ0132.6-0804  &  59127 (2020-10-05)  &  23 $\pm$ 3  &  59143 (2020-10-21)  &  88 $\pm$ 2  &  16  &  -0.44 $\pm$ 0.03 \\
RXJ0132.6-0804  &  59248 (2021-02-03)  &  25 $\pm$ 3  &  59183 (2020-11-30)  &  82 $\pm$ 3  &  65  &  -0.39 $\pm$ 0.02 \\
J0439+0520  &  56201 (2012-10-01)  &  18 $\pm$ 5  &  56200 (2012-09-30)  &  318 $\pm$ 8  &  1  &  -0.91 $\pm$ 0.06 \\
J0439+0520  &  56496 (2013-07-23)  &  41 $\pm$ 13  &  56494 (2013-07-20)  &  326 $\pm$ 40  &  2  &  -0.65 $\pm$ 0.07 \\
J0439+0520  &  58568 (2019-03-26)  &  19 $\pm$ 4  &  58565 (2019-03-22)  &  342 $\pm$ 4  &  3  &  -0.92 $\pm$ 0.06 \\
J0439+0520  &  58750 (2019-09-24)  &  29 $\pm$ 2  &  58738 (2019-09-11)  &  346 $\pm$ 5  &  12  &  -0.78 $\pm$ 0.02 \\
J0439+0520  &  58920 (2020-03-12)  &  26 $\pm$ 2  &  58926 (2020-03-18)  &  336 $\pm$ 5  &  6  &  -0.81 $\pm$ 0.02 \\
J0439+0520  &  59096 (2020-09-04)  &  37 $\pm$ 2  &  59097 (2020-09-04)  &  336 $\pm$ 5  &  1  &  -0.69 $\pm$ 0.01 \\
J0439+0520  &  59247 (2021-02-02)  &  36 $\pm$ 2  &  59193 (2020-12-10)  &  321 $\pm$ 4  &  54  &  -0.7 $\pm$ 0.01 \\
A646  &  55984 (2012-02-27)  &  16 $\pm$ 6  &  56323 (2013-01-31)  &  44 $\pm$ 12  &  339  &  -0.35 $\pm$ 0.07 \\
A646  &  57047 (2015-01-25)  &  12 $\pm$ 4  &  57055 (2015-02-02)  &  59 $\pm$ 9  &  8  &  -0.52 $\pm$ 0.05 \\
A646  &  58576 (2019-04-03)  &  10 $\pm$ 2  &  58573 (2019-03-30)  &  65 $\pm$ 4  &  3  &  -0.57 $\pm$ 0.05 \\
A646  &  58756 (2019-09-30)  &  8 $\pm$ 3  &  58738 (2019-09-11)  &  57 $\pm$ 2  &  18  &  -0.67 $\pm$ 0.1 \\
A646  &  58917 (2020-03-09)  &  9 $\pm$ 2  &  58930 (2020-03-22)  &  55 $\pm$ 1  &  13  &  -0.58 $\pm$ 0.06 \\
A646  &  59120 (2020-09-28)  &  12 $\pm$ 3  &  59109 (2020-09-16)  &  53 $\pm$ 2  &  11  &  -0.46 $\pm$ 0.05 \\
4C55.16  &  55984 (2012-02-27)  &  27 $\pm$ 6  &  56342 (2013-02-19)  &  486 $\pm$ 37  &  358  &  -0.94 $\pm$ 0.05 \\
4C55.16  &  58567 (2019-03-25)  &  21 $\pm$ 4  &  58565 (2019-03-23)  &  538 $\pm$ 19  &  2  &  -1.02 $\pm$ 0.05 \\
4C55.16  &  58750 (2019-09-24)  &  31 $\pm$ 3  &  58739 (2019-09-12)  &  546 $\pm$ 13  &  11  &  -0.9 $\pm$ 0.01 \\
4C55.16  &  58917 (2020-03-09)  &  24 $\pm$ 2  &  58920 (2020-03-12)  &  510 $\pm$ 20  &  3  &  -0.98 $\pm$ 0.01 \\
4C55.16  &  59117 (2020-09-25)  &  31 $\pm$ 3  &  59100 (2020-09-07)  &  486 $\pm$ 13  &  17  &  -0.87 $\pm$ 0.01 \\
A1348  &  56288 (2012-12-27)  &  14 $\pm$ 5  &  56336 (2013-02-13)  &  61 $\pm$ 4  &  48  &  -0.39 $\pm$ 0.08 \\
A1348  &  56479 (2013-07-06)  &  10 $\pm$ 5  &  56465 (2013-06-22)  &  52 $\pm$ 4  &  14  &  -0.5 $\pm$ 0.09 \\
A1348  &  58583 (2019-04-10)  &  3 $\pm$ 2  &  58593 (2019-04-20)  &  40 $\pm$ 2  &  10  &  -0.86 $\pm$ 0.11 \\
A1348  &  58813 (2019-11-26)  &  2 $\pm$ 3  &  58811 (2019-11-23)  &  47 $\pm$ 3  &  2  &  -1.05 $\pm$ 0.32 \\
A1348  &  58917 (2020-03-09)  &  2 $\pm$ 2  &  58917 (2020-03-09)  &  43 $\pm$ 4  &  0  &  -0.97 $\pm$ 0.21 \\
3C264  &  58583 (2019-04-10)  &  56 $\pm$ 3  &  58582 (2019-04-09)  &  142 $\pm$ 9  &  1  &  -0.31 $\pm$ -0.01 \\
3C264  &  58793 (2019-11-06)  &  65 $\pm$ 3  &  58793 (2019-11-05)  &  115 $\pm$ 23  &  0  &  -0.18 $\pm$ -0.03 \\
3C264  &  58914 (2020-03-06)  &  58 $\pm$ 2  &  58913 (2020-03-05)  &  128 $\pm$ 9  &  1  &  -0.25 $\pm$ -0.01 \\
3C264  &  59194 (2020-12-11)  &  73 $\pm$ 2  &  59192 (2020-12-08)  &  140 $\pm$ 10  &  2  &  -0.22 $\pm$ -0.01 \\
A1644  &  58600 (2019-04-27)  &  36 $\pm$ 3  &  58597 (2019-04-24)  &  166 $\pm$ 3  &  3  &  -0.49 $\pm$ 0.02 \\
A1644  &  58803 (2019-11-16)  &  34 $\pm$ 3  &  58854 (2020-01-05)  &  215 $\pm$ 3  &  51  &  -0.57 $\pm$ 0.02 \\
A1644  &  58917 (2020-03-09)  &  42 $\pm$ 2  &  58926 (2020-03-18)  &  214 $\pm$ 9  &  9  &  -0.51 $\pm$ 0.01 \\
A1644  &  59184 (2020-12-01)  &  51 $\pm$ 2  &  59182 (2020-11-28)  &  200 $\pm$ 3  &  2  &  -0.42 $\pm$ 0.01 \\
J1350+0940  &  58980 (2020-05-11)  &  14 $\pm$ 5  &  58979 (2020-05-10)  &  135 $\pm$ 3  &  1  &  -0.72 $\pm$ 0.09 \\
J1350+0940  &  58605 (2019-05-02)  &  12 $\pm$ 3  &  58603 (2019-04-30)  &  133 $\pm$ 3  &  2  &  -0.77 $\pm$ 0.05 \\
J1350+0940  &  58847 (2019-12-30)  &  14 $\pm$ 2  &  58854 (2020-01-05)  &  135 $\pm$ 2  &  7  &  -0.71 $\pm$ 0.04 \\
J1350+0940  &  58994 (2020-05-25)  &  12 $\pm$ 3  &  58994 (2020-05-25)  &  134 $\pm$ 3  &  0  &  -0.75 $\pm$ 0.05 \\
J1350+0940  &  59214 (2020-12-31)  &  13 $\pm$ 1  &  59181 (2020-11-27)  &  131 $\pm$ 2  &  33  &  -0.74 $\pm$ 0.03 \\
J1459-1810  &  56931 (2014-10-01)  &  13 $\pm$ 6  &  56931 (2014-09-30)  &  142 $\pm$ 11  &  0  &  -0.74 $\pm$ 0.1 \\
J1459-1810  &  58599 (2019-04-26)  &  9 $\pm$ 2  &  58603 (2019-04-30)  &  133 $\pm$ 3  &  4  &  -0.85 $\pm$ 0.07 \\
J1459-1810  &  58879 (2020-01-31)  &  15 $\pm$ 3  &  58877 (2020-01-28)  &  135 $\pm$ 3  &  2  &  -0.72 $\pm$ 0.05 \\
J1459-1810  &  58991 (2020-05-22)  &  11 $\pm$ 3  &  58995 (2020-05-26)  &  141 $\pm$ 3  &  4  &  -0.79 $\pm$ 0.06 \\
A2052  &  56320 (2013-01-28)  &  22 $\pm$ 7  &  56320 (2013-01-27)  &  110 $\pm$ 10  &  0  &  -0.43 $\pm$ 0.07 \\
A2052  &  56476 (2013-07-03)  &  22 $\pm$ 6  &  56472 (2013-06-29)  &  108 $\pm$ 5  &  4  &  -0.53 $\pm$ 0.06 \\
A2052  &  58599 (2019-04-26)  &  27 $\pm$ 2  &  58594 (2019-04-21)  &  90 $\pm$ 6  &  5  &  -0.34 $\pm$ 0.01 \\
A2052  &  58871 (2020-01-23)  &  26 $\pm$ 2  &  58873 (2020-01-24)  &  71 $\pm$ 5  &  2  &  -0.35 $\pm$ 0.01 \\
A2052  &  58994 (2020-05-25)  &  23 $\pm$ 3  &  58985 (2020-05-16)  &  69 $\pm$ 4  &  9  &  -0.39 $\pm$ 0.02 \\
A2052  &  59247 (2021-02-02)  &  22 $\pm$ 3  &  59193 (2020-12-09)  &  69 $\pm$ 3  &  54  &  -0.34 $\pm$ 0.02 \\
A2055  &  58599 (2019-04-26)  &  15 $\pm$ 2  &  58603 (2019-04-30)  &  32 $\pm$ 3  &  4  &  -0.25 $\pm$ 0.02 \\
A2055  &  58917 (2020-03-09)  &  18 $\pm$ 2  &  58909 (2020-02-29)  &  27 $\pm$ 3  &  8  &  -0.16 $\pm$ 0.01 \\
A2055  &  59017 (2020-06-17)  &  17 $\pm$ 2  &  59015 (2020-06-15)  &  30 $\pm$ 2  &  2  &  -0.25 $\pm$ 0.02 \\
J1558-1409  &  56032 (2012-04-15)  &  33 $\pm$ 4  &  56034 (2012-04-17)  &  141 $\pm$ 13  &  2  &  -0.47 $\pm$ 0.01 \\
J1558-1409  &  56476 (2013-07-03)  &  28 $\pm$ 6  &  56472 (2013-06-29)  &  129 $\pm$ 13  &  4  &  -0.49 $\pm$ 0.04 \\
J1558-1409  &  57047 (2015-01-25)  &  31 $\pm$ 4  &  57045 (2015-01-22)  &  121 $\pm$ 9  &  2  &  -0.48 $\pm$ 0.01 \\
J1558-1409  &  58599 (2019-04-26)  &  32 $\pm$ 2  &  58596 (2019-04-23)  &  142 $\pm$ 4  &  3  &  -0.51 $\pm$ 0.01 \\
J1558-1409  &  58879 (2020-01-31)  &  24 $\pm$ 3  &  58873 (2020-01-24)  &  151 $\pm$ 5  &  6  &  -0.56 $\pm$ 0.03 \\
J1558-1409  &  58991 (2020-05-22)  &  24 $\pm$ 2  &  58995 (2020-05-26)  &  139 $\pm$ 3  &  4  &  -0.55 $\pm$ 0.02 \\
\hline
	\end{tabular}
	\label{tab:SCUBA2_obs1}
\end{table*}

\begin{table*}
	\centering
	\begin{tabular}{lcccccc}
\hline
Source & SCUBA2 Obs. & SCUBA2 flux density & OVRO Obs.& OVRO flux density & Time Difference & $\alpha$ \\
 & mjd (ISO) & / mJy & mjd (ISO) & / mJy & / days & \\
\hline
J1558-1409  &  59247 (2021-02-02)  &  25 $\pm$ 3  &  59189 (2020-12-05)  &  124 $\pm$ 3  &  58  &  -0.52 $\pm$ 0.02 \\
J1603+1554  &  58547 (2019-03-05)  &  30 $\pm$ 3  &  58540 (2019-02-26)  &  292 $\pm$ 5  &  7  &  -0.71 $\pm$ 0.02 \\
J1603+1554  &  58757 (2019-10-01)  &  27 $\pm$ 3  &  58736 (2019-09-10)  &  310 $\pm$ 6  &  21  &  -0.77 $\pm$ 0.03 \\
J1603+1554  &  58914 (2020-03-06)  &  31 $\pm$ 3  &  58917 (2020-03-08)  &  325 $\pm$ 4  &  3  &  -0.74 $\pm$ 0.02 \\
J1603+1554  &  59306 (2021-04-02)  &  40 $\pm$ 3  &  59189 (2020-12-05)  &  318 $\pm$ 5  &  117  &  -0.66 $\pm$ 0.02 \\
J1727+5510  &  56200 (2012-09-30)  &  31 $\pm$ 4  &  56197 (2012-09-27)  &  251 $\pm$ 25  &  3  &  -0.64 $\pm$ 0.02 \\
J1727+5510  &  56476 (2013-07-03)  &  29 $\pm$ 6  &  56478 (2013-07-05)  &  260 $\pm$ 8  &  2  &  -0.7 $\pm$ 0.04 \\
J1727+5510  &  56480 (2013-07-07)  &  22 $\pm$ 4  &  56478 (2013-07-05)  &  260 $\pm$ 8  &  2  &  -0.79 $\pm$ 0.04 \\
J1727+5510  &  58547 (2019-03-05)  &  34 $\pm$ 3  &  58555 (2019-03-12)  &  309 $\pm$ 5  &  8  &  -0.7 $\pm$ 0.02 \\
J1727+5510  &  58756 (2019-09-30)  &  35 $\pm$ 3  &  58747 (2019-09-21)  &  331 $\pm$ 4  &  9  &  -0.71 $\pm$ 0.02 \\
J1727+5510  &  58908 (2020-02-29)  &  41 $\pm$ 2  &  58909 (2020-02-29)  &  335 $\pm$ 4  &  1  &  -0.66 $\pm$ 0.01 \\
J1727+5510  &  59292 (2021-03-19)  &  41 $\pm$ 2  &  59191 (2020-12-07)  &  329 $\pm$ 6  &  101  &  -0.66 $\pm$ 0.01 \\
Z8276  &  56200 (2012-09-30)  &  21 $\pm$ 4  &  56327 (2013-02-03)  &  105 $\pm$ 4  &  127  &  -0.44 $\pm$ 0.05 \\
Z8276  &  56476 (2013-07-03)  &  24 $\pm$ 5  &  56477 (2013-07-04)  &  88 $\pm$ 7  &  1  &  -0.44 $\pm$ 0.04 \\
Z8276  &  58547 (2019-03-05)  &  19 $\pm$ 3  &  58557 (2019-03-15)  &  73 $\pm$ 2  &  10  &  -0.42 $\pm$ 0.04 \\
Z8276  &  58756 (2019-09-30)  &  24 $\pm$ 3  &  58739 (2019-09-12)  &  78 $\pm$ 2  &  17  &  -0.37 $\pm$ 0.02 \\
Z8276  &  58918 (2020-03-10)  &  33 $\pm$ 2  &  58917 (2020-03-08)  &  76 $\pm$ 2  &  1  &  -0.26 $\pm$ 0.01 \\
Z8276  &  59292 (2021-03-19)  &  28 $\pm$ 2  &  59184 (2020-11-30)  &  77 $\pm$ 2  &  108  &  -0.31 $\pm$ 0.01 \\
RGBJ1745+398  &  58564 (2019-03-22)  &  13 $\pm$ 2  &  58567 (2019-03-24)  &  46 $\pm$ 2  &  3  &  -0.36 $\pm$ 0.02 \\
RGBJ1745+398  &  58751 (2019-09-25)  &  13 $\pm$ 3  &  58735 (2019-09-09)  &  38 $\pm$ 2  &  16  &  -0.34 $\pm$ 0.04 \\
RGBJ1745+398  &  58914 (2020-03-06)  &  16 $\pm$ 2  &  58917 (2020-03-08)  &  35 $\pm$ 2  &  3  &  -0.24 $\pm$ 0.02 \\
RGBJ1745+398  &  59214 (2020-12-31)  &  12 $\pm$ 1  &  59173 (2020-11-19)  &  31 $\pm$ 3  &  41  &  -0.34 $\pm$ 0.01 \\
A2390  &  56406 (2013-04-24)  &  11 $\pm$ 4  &  56405 (2013-04-22)  &  48 $\pm$ 4  &  1  &  -0.48 $\pm$ 0.03 \\
A2390  &  56433 (2013-05-21)  &  12 $\pm$ 6  &  56438 (2013-05-25)  &  43 $\pm$ 8  &  5  &  -0.42 $\pm$ 0.08 \\
A2390  &  58599 (2019-04-26)  &  7 $\pm$ 3  &  58599 (2019-04-25)  &  47 $\pm$ 2  &  0  &  -0.61 $\pm$ 0.08 \\
A2390  &  58752 (2019-09-26)  &  6 $\pm$ 3  &  58734 (2019-09-08)  &  54 $\pm$ 3  &  18  &  -0.69 $\pm$ 0.11 \\
A2390  &  59018 (2020-06-18)  &  9 $\pm$ 2  &  59018 (2020-06-18)  &  52 $\pm$ 2  &  0  &  -0.57 $\pm$ 0.06 \\
A2390  &  59214 (2020-12-31)  &  7 $\pm$ 1  &  59197 (2020-12-14)  &  53 $\pm$ 3  &  17  &  -0.63 $\pm$ 0.04 \\
A2390  &  59373 (2021-06-08)  &  11 $\pm$ 3  &  59197 (2020-12-14)  &  53 $\pm$ 3  &  176  &  -0.51 $\pm$ 0.06 \\
A2415  &  56201 (2012-10-01)  &  18 $\pm$ 6  &  56328 (2013-02-04)  &  107 $\pm$ 4  &  127  &  -0.57 $\pm$ 0.08 \\
A2415  &  56475 (2013-07-02)  &  18 $\pm$ 9  &  56479 (2013-07-06)  &  123 $\pm$ 4  &  4  &  -0.6 $\pm$ 0.11 \\
A2415  &  58567 (2019-03-25)  &  20 $\pm$ 2  &  58565 (2019-03-22)  &  159 $\pm$ 3  &  2  &  -0.67 $\pm$ 0.02 \\
A2415  &  58750 (2019-09-24)  &  19 $\pm$ 2  &  58736 (2019-09-10)  &  122 $\pm$ 2  &  14  &  -0.59 $\pm$ 0.03 \\
A2415  &  59018 (2020-06-18)  &  21 $\pm$ 2  &  59019 (2020-06-18)  &  107 $\pm$ 2  &  1  &  -0.54 $\pm$ 0.03 \\
A2415  &  59127 (2020-10-05)  &  16 $\pm$ 3  &  59144 (2020-10-22)  &  107 $\pm$ 2  &  17  &  -0.61 $\pm$ 0.05 \\
A2415  &  59322 (2021-04-18)  &  15 $\pm$ 2  &  59192 (2020-12-09)  &  106 $\pm$ 3  &  130  &  -0.62 $\pm$ 0.04 \\
A2627  &  56224 (2012-10-24)  &  21 $\pm$ 5  &  56349 (2013-02-25)  &  35 $\pm$ 7  &  125  &  -0.21 $\pm$ 0.04 \\
A2627  &  56475 (2013-07-02)  &  23 $\pm$ 7  &  56475 (2013-07-01)  &  43 $\pm$ 8  &  0  &  -0.23 $\pm$ 0.02 \\
A2627  &  58615 (2019-05-12)  &  16 $\pm$ 2  &  58619 (2019-05-15)  &  37 $\pm$ 3  &  4  &  -0.27 $\pm$ 0.02 \\
A2627  &  58751 (2019-09-25)  &  17 $\pm$ 3  &  58733 (2019-09-07)  &  45 $\pm$ 3  &  18  &  -0.32 $\pm$ 0.03 \\
A2627  &  59029 (2020-06-29)  &  13 $\pm$ 3  &  59028 (2020-06-28)  &  49 $\pm$ 2  &  1  &  -0.4 $\pm$ 0.05 \\
A2627  &  59230 (2021-01-16)  &  12 $\pm$ 2  &  59191 (2020-12-08)  &  41 $\pm$ 3  &  39  &  -0.41 $\pm$ 0.04 \\
RXJ2341.1+0018  &  56170 (2012-08-31)  &  27 $\pm$ 4  &  56329 (2013-02-05)  &  106 $\pm$ 4  &  159  &  -0.39 $\pm$ 0.03 \\
RXJ2341.1+0018  &  56189 (2012-09-19)  &  23 $\pm$ 4  &  56329 (2013-02-05)  &  106 $\pm$ 4  &  140  &  -0.44 $\pm$ 0.04 \\
RXJ2341.1+0018  &  56476 (2013-07-03)  &  30 $\pm$ 5  &  56481 (2013-07-07)  &  104 $\pm$ 6  &  5  &  -0.39 $\pm$ 0.03 \\
RXJ2341.1+0018  &  58598 (2019-04-25)  &  18 $\pm$ 2  &  58599 (2019-04-25)  &  98 $\pm$ 2  &  1  &  -0.52 $\pm$ 0.03 \\
RXJ2341.1+0018  &  58751 (2019-09-25)  &  19 $\pm$ 2  &  58738 (2019-09-12)  &  96 $\pm$ 3  &  13  &  -0.51 $\pm$ 0.02 \\
\hline
	\end{tabular}
	\label{tab:SCUBA2_obs2}
\end{table*}

\section{ALMA Data}
\label{sec:ALMA_data}
\begin{table*}
    \caption{ALMA flux densities, and the spectral indices calculated with the nearest 15 GHz OVRO 40m observations.
    *The observation of A2052 was completed on two separate dates (mjd 57611 and 57623), and we use the central date to find the nearest OVRO measurement for calculating the spectral index.}
	\centering
	\begin{tabular}{lccccccc}
\hline
Source & ALMA Obs. & ALMA Freq. & ALMA flux den. & OVRO Obs. & OVRO flux den. & Time diff. & $\alpha$ \\
 & mjd (ISO) & / GHz & / mJy & mjd (ISO) & / mJy & / days & \\
\hline
RXJ0132.6-0804 & 58134.0 (2018-01-16) & 92.7 & 31.6 $\pm$ 0.6 & 58132 (2018-01-14) & 101 $\pm$ 2 & 1 & -0.65 $\pm$ -0.03 \\
J0439+0520 & 58119.0 (2018-01-01) & 100.1 & 34.6 $\pm$ 1.0 & 58122 (2018-01-04) & 341 $\pm$ 5 & 3 & -1.22 $\pm$ 0.01 \\
A1644 & 58351.0 (2018-08-21) & 103.1 & 29.2 $\pm$ 1.0 & 58449 (2018-11-27) & 164 $\pm$ 2 & 98 & -0.95 $\pm$ 0.03 \\
J1350+0940 & 58377.0 (2018-09-16) & 94.8 & 10.2 $\pm$ 0.4 & 58408 (2018-10-17) & 129 $\pm$ 2 & 31 & -1.39 $\pm$ 0.01 \\
A2052 & 57617 (2016-08-17)* & 229.5 & 32.5 $\pm$ 0.1 & 57612 (2016-08-12) & 93 $\pm$ 3 & 4 & -0.40 $\pm$ -0.04 \\
A2390 & 58125.0 (2018-01-07) & 98.3 & 7.3 $\pm$ 0.4 & 58131 (2018-01-13) & 44 $\pm$ 2 & 6 & -1.0 $\pm$ 0.01 \\
A2415 & 58141.0 (2018-01-23) & 102.0 & 19.7 $\pm$ 0.3 & 58141 (2018-01-23) & 152 $\pm$ 2 & 0 & -1.09 $\pm$ -0.01 \\
\hline
	\end{tabular}
	\label{tab:ALMA_obs}
\end{table*}


\bsp	
\label{lastpage}
\end{document}